\documentclass[10pt,twocolumn,showpacs,preprintnumbers,amsmath,amssymb,aps,prd,nofootinbib,superscriptaddress,longbibliography]{revtex4-1}
\usepackage{wasysym}
\usepackage[utf8]{inputenc}
\usepackage[english]{babel}
\usepackage{epsfig}
\usepackage{subfigure}
\usepackage{bm}
\usepackage{amsfonts}
\usepackage{dcolumn}
\usepackage{hyperref}

\hypersetup{colorlinks=true, linkcolor=blue, citecolor=green}

\usepackage{dcolumn}
\usepackage{bm}
\usepackage{ifpdf}
\usepackage{bm}
\usepackage{xcolor,color,graphicx,graphics}
\usepackage[OT1]{fontenc}
\usepackage{latexsym,amssymb,amsmath,amsfonts}
\usepackage{makeidx}
\usepackage{epstopdf}
\usepackage{mathrsfs}
\hypersetup{colorlinks=true, linkcolor=blue, citecolor=green}
\usepackage{enumerate}
 \usepackage{multirow}

\begin{document}

\title{Roche limit and stellar disruption in the Simpson--Visser spacetime}


\author{Marcos V. de S. Silva}
\email{marcos.sousa@uva.es}
\affiliation{Department of Theoretical Physics, Atomic and Optics, Campus Miguel Delibes, \\ University of Valladolid UVA, Paseo Bel\'en, 7, 47011 - Valladolid, Spain}
     

\begin{abstract}
Due to the tidal forces that a black hole can produce, certain types of compact objects may undergo disruption as they approach the black hole. This disruption point is known as the Roche limit (or Roche radius). In this work, we studied the tidal forces arising from the presence of the Simpson--Visser black bounce. We analyzed the tidal forces both for a static observer and for a radially infalling observer and showed that differences arise depending on the choice of observer. We used the tidal forces together with the stellar binding forces to determine the Roche radius for neutron stars, white dwarfs, and Sun-like stars, and to investigate how the Simpson--Visser regularization affects the tidal disruption of these astrophysical objects. We also examined whether, for astrophysical black holes such as M87* and Sgr~A*, these stellar disruption processes occur inside or outside the event horizon, and thus whether they are observable. To provide a more realistic dynamical description, we implement the Affine Model to evaluate the tidal deformation of neutron stars, white dwarfs, and main-sequence stars, assessing how the regularized geometry and the black hole mass govern the evolution of the stellar axes.

\end{abstract}

\date{\today}

\maketitle

\section{Introduction}
The detection of black hole shadows represents an important point in observational astrophysics and provides a direct probe of the strong-field regime of gravity \cite{EventHorizonTelescope:2022wkp,EventHorizonTelescope:2019dse}. Using very-long-baseline interferometry at millimeter wavelengths, the Event Horizon Telescope (EHT) collaboration achieved the first images of the shadow of the supermassive black hole as the M87* and Sgr~A* \cite{EventHorizonTelescope:2022wkp,EventHorizonTelescope:2022apq,EventHorizonTelescope:2019dse,EventHorizonTelescope:2019uob,EventHorizonTelescope:2019jan}. These observations revealed a dark central region surrounded by a bright emission ring, consistent with theoretical predictions for photon capture by a black hole and gravitational lensing near the event horizon. The measured size and shape of the shadows agree with the expectations of general relativity, allowing stringent tests of Einstein's theory in the strong-gravity regime and placing constraints on alternative models of compact objects \cite{Vagnozzi:2022moj}. In addition, shadow observations provide valuable information about black hole masses, distance from the black hole, spins, and the properties of the surrounding accretion flows \cite{Vagnozzi:2022moj}.

The detection of gravitational waves and black hole shadows has strongly reinforced the theoretical prediction of black holes, since the astrophysical observations are in remarkable agreement with the theoretical expectations for these objects \cite{Barack:2018yly,Cardoso:2017cqb,LISA:2022kgy,Cunha:2018acu,Herdeiro:2021lwl,Sengo:2024pwk,Lima:2021las,Bambi:2019tjh,Khodadi:2024ubi}. Despite this compelling evidence, neither shadow observations nor gravitational-wave detections constitute definitive proof of the existence of black holes, as other theoretical models can mimic similar signatures in certain regimes \cite{Cardoso:2016rao,Cunha:2018gql}. Alternative compact objects, such as gravastars, boson stars, or wormholes, may reproduce features of the gravitational-wave signals or the optical appearance of black hole shadows \cite{Herdeiro:2021lwl,Rosa:2024bqv,Rosa:2023qcv,Cardoso:2016rao,Lima:2021las,Guerrero:2021ues,Guerrero:2022qkh}. Therefore, although current observations provide strong support for the black hole paradigm, a complete and unambiguous identification still requires further observational tests capable of discriminating between black holes and these exotic alternatives.

An alternative to standard black holes is provided by regular black holes \cite{Ansoldi:2008jw}. These compact objects possess an event horizon, as in the usual black hole case, but differ by not containing singularities. In the context of general relativity, regular black holes are not vacuum solutions and require nonstandard matter sources that violate the energy conditions in order to be sustained, such as nonlinear electrodynamics, scalar fields, dark matter, and other fields \cite{Bronnikov:2000vy,Bronnikov:2005gm,Bronnikov:2022ofk,Hayward:2005gi,Konoplya:2025ect}. The first regular black hole solution was proposed by Bardeen and later interpreted by Ayón-Beato and García as a solution of Einstein's equations sourced by nonlinear electrodynamics \cite{Ayon-Beato:2000mjt,Rodrigues:2018bdc}. Since then, many regular black hole models have been proposed and extensively studied in the literature \cite{Rodrigues:2017yry,Ayon-Beato:1998hmi,Ayon-Beato:1999kuh,Dymnikova:2004zc,Balart:2014cga,Junior:2020zdt,Rodrigues:2019xrc,Rodrigues:2022qdp,Rodrigues:2020pem,deSousaSilva:2018kkt,Lan:2023cvz,Bueno:2024dgm,Fontana:2023zqz,Estrada:2024uuu,Bueno:2025dqk,Fernandes:2025eoc,Hennigar:2025yqm,dePaula:2023ozi,dePaula:2025emo,dePaula:2025kif,dePaula:2024yzy,Dolan:2024qqr,dePaula:2024xnd,Balart:2023odm,Bonanno:2023rzk,Bueno:2024eig}.

Another class of regular solutions is provided by the so-called black bounces. These models, originally proposed by Hochberg and Visser and later studied in greater detail by Simpson and Visser \cite{Visser:1997yn,Simpson:2018tsi}, are characterized by the presence of a wormhole throat hidden behind an event horizon. One of the most prominent black bounce models is the Simpson--Visser solution \cite{Simpson:2018tsi}. This model arises from a regularization procedure applied to the Schwarzschild metric, in which the central singularity and its neighborhood are removed and replaced by a wormhole throat \cite{Franzin:2021vnj,Lobo:2020ffi}. Within general relativity, this solution was initially obtained by considering a minimal coupling between a phantom scalar field and nonlinear electrodynamics \cite{Canate:2022gpy,Bronnikov:2021uta,Rodrigues:2023vtm}. It is also possible to reproduce this solution by considering linear electrodynamics with a nonminimal coupling to the scalar field, or within alternative theories of gravity \cite{Cordeiro:2025ivw,Fabris:2023opv,Rodrigues:2024iao,Silva:2025fqj}. Since the original work of Simpson and Visser, many other black bounce models have been proposed and investigated in a variety of contexts, including modified gravity theories, extra dimensions, rotating solutions, $2+1$-dimensional spacetimes, and cylindrical configurations \cite{Rodrigues:2022mdm,Rodrigues:2022rfj,Bronnikov:2023aya,Alencar:2024yvh,Pereira:2024rtv,Alencar:2024nxi,Pereira:2025xnw,Silva:2025eip,Alencar:2025jvl,Muniz:2024wiv,Crispim:2024nou,Crispim:2024yjz,Lima:2022pvc,Lima:2023arg,Junior:2024cbb,Boos:2020qgg,Boos:2021kqe,Silva:2026gvg,Kirezli:2025cjy,Rodrigues:2025plw,Lessa:2024erf,Junior:2024vrv,Atazadeh:2023wdw,Junior:2023qaq,Alencar:2026qeb,Cordeiro:2025ivw,Maeda:2025bnb,Junior:2025sjr}.

One of the most remarkable features of the Simpson--Visser solution is its ability to mimic the Schwarzschild spacetime \cite{Lima:2021las,Guerrero:2021ues,Silva:2024fpn,Franzin:2023slm}. Although it possesses a different photon ring structure, corresponding to unstable circular photon orbits, the shadow produced by the Simpson--Visser model is very similar to that of Schwarzschild, both in the case of a celestial sphere and in its optical appearance. Several works have explored the properties of this spacetime and its similarities with singular black hole solutions \cite{Pereira:2025fvg,Nascimento:2020ime,Furtado:2025zva}. Other black bounce models arising in different theoretical frameworks have also had their physical and observational properties extensively analyzed in the literature \cite{Soares:2025hpy,Soares:2024rhp,Soares:2023uup,Calza:2024xdh,Calza:2025yfm,Duran-Cabaces:2025sly,Borissova:2025msp,Yang:2022ryf,Siqueira:2026uzu,Nascimento:2025mtr,Moreira:2025lip,Shuai:2025rvr,Bragado:2025jrg,Li:2025yoz,Santos:2025xbk,Eiroa:2025mws,Olmo:2023lil}.

One of the most effective ways to analyze the physical properties of a black hole is through the behavior of matter and fields in its vicinity. A prominent example is the optical appearance of black holes \cite{Guerrero:2021ues,Guerrero:2022qkh}, which arises from both the distribution of surrounding matter, such as accretion disks, and the motion of photons emitted near the horizon. Another example is the dynamics of planets and stars orbiting supermassive black holes \cite{GRAVITY:2018ofz,GRAVITY:2020gka,Genzel:2010zy,Ishibashi:2024wly}. An important probe is provided by the tidal forces. Tidal forces arise from the inhomogeneity of the gravitational field, which generates a gradient of gravitational acceleration across an extended body \cite{Hobson:2006se}. This differential force produces a stretching effect and, if sufficiently strong, can overcome the internal binding forces of the object, leading to tidal disruption \cite{Gezari:2021bmb,Joshi:2025nzb,Komossa:2015qya}. Such events commonly occur when astrophysical objects, such as stars, move in the vicinity of a black hole. The radius at which disruption takes place is known as the Roche limit or Roche radius \cite{Evans:1989qe,Junior:2026upy}. Since different black hole models generate different tidal force profiles, it is expected that their corresponding Roche limits may also be modified in a way that could, in principle, be observed astrophysically \cite{LimaJunior:2022gko}.

For the Schwarzschild black hole, the tidal forces diverge near the central singularity \cite{dInverno:1992gxs,Hobson:2006se}; however, this divergence is not observable because it is hidden behind the event horizon. In the Reissner--Nordström case, a particularly interesting phenomenon occurs: the tidal forces undergo a sign inversion, so that the body experiences compression instead of the usual stretching \cite{Crispino:2016pnv,Sharif:2018gzj,LimaJunior:2022gko}. Not all black hole models exhibit divergent tidal forces; regular solutions such as the Bardeen black hole, among others, display finite tidal forces everywhere \cite{LimaJunior:2022gko,Lima:2020wcb,Crispim:2025cql,Sharif:2018gaj}. In this way, tidal forces provide a powerful and insightful tool to investigate the behavior of matter in the vicinity of black holes and to discriminate between different theoretical models \cite{Albacete:2024qja,Silva:2025hwl,Arora:2023ltv}. While tidal forces have been extensively analyzed in the context of black holes, the Simpson--Visser spacetime introduces a fundamentally different physical paradigm. Standard regular black holes typically replace the central singularity with a de Sitter-like repulsive core at $r=0$. In contrast, the Simpson--Visser geometry is a black bounce model that modifies the topology of the spacetime, replacing the central singularity with a throat at $r=a$ \cite{Uniyal:2025sdr}. A primary novelty of this work is to explore how this topological shift imprints unique signatures on the tidal tensor. Unlike standard regular geometries, the presence of the Simpson--Visser throat can lead to an extreme modification of the tidal field, including a potential sign inversion of the tidal forces, directly affecting the Roche limit and stellar disruption in ways not captured by previous literature. To some extent, the Roche limit itself has not been explored in the literature as extensively as the topic of tidal forces.

The structure of this paper is organized as follows. In Section~\ref{SEC:spacetime}, we present the spacetime under consideration and discuss some of its main properties, as well as the radial geodesics in this geometry. In Section~\ref{SEC:TF}, we derive the tidal forces for both a static observer and a radially infalling observer. In Section~\ref{SEC:Roche}, we use the tidal forces to compute the Roche limit for this spacetime. Finally, the discussions, conclusions, and perspectives for future work are presented in Section~\ref{Sec:Conclusion}.

\section{Spacetime and Geodesics}\label{SEC:spacetime}
Black bounces can be characterized as spacetimes in which the throat of a wormhole is hidden by the presence of horizons. The most well-known black bounce model is described by the Simpson--Visser spacetime, whose line element is usually written as \cite{Simpson:2018tsi}
\begin{equation}\label{linex}
\begin{split}
    ds^2&=-A(x)dt^2+\frac{1}{A(x)}dx^2+\Sigma(x)^2(d\theta^2+\sin^2\theta d\phi^2),\\
    A(x)&=1-\frac{2M}{\Sigma(x)}, \qquad \Sigma(x)=\sqrt{x^2+a^2}.
\end{split}
\end{equation}
This spacetime features a wormhole throat located at $x_t=0$ and an event horizon located at $x_+ = \sqrt{4M^2-a^2}$. It also possesses several other interesting properties. One example is the fact that, although the radius of the unstable photon orbit is given by $x_{ph}=\sqrt{9M^2-a^2}$ for $a\leq 3M$, the radius of the apparent shadow of this spacetime is
$x_S = 3\sqrt{3}\,M$ \cite{Vagnozzi:2022moj,Silva:2024fpn,Lima:2020auu,LimaJunior:2022zvu}, which coincides with the Schwarzschild result. Therefore, from the viewpoint of optical appearance and black hole shadow, the Simpson--Visser solution mimics the Schwarzschild black hole with remarkable efficiency \cite{Lima:2021las,Guerrero:2021ues,Franzin:2023slm}.

Alternatively, by performing the coordinate transformation $r=\Sigma(x)$, we can rewrite the line element describing the Simpson--Visser spacetime as \cite{Franzin:2023slm, Silva:2025eip,Crispim:2025cql}
\begin{equation}\label{line}
\begin{split}
    ds^2=-A(r)dt^2+\frac{1}{B(r)}dr^2+r^2(d\theta^2+\sin^2\theta d\phi^2),\\
    A(r)=1-\frac{2M}{r}, \quad B(r)=\left(1-\frac{2M}{r}\right)\left(1-\frac{a^2}{r^2}\right).
\end{split}
\end{equation}
In this coordinate system, the wormhole throat is located at $r_t=a$, while the event horizon radius is located at $r_+=2M$. It thus becomes clear that, if $a>2M$, the throat radius exceeds the event horizon radius, so that the event horizon is no longer accessible and the spacetime describes a two-way traversable wormhole. In the case $a=2M$, the horizon coincides with the wormhole throat, $r_t=r_+$, and the spacetime describes a one-way traversable wormhole. Thus, when the Simpson--Visser spacetime is written in terms of the coordinate $r$, we do not need to worry about the location of the event horizon radius for each value of $a$.

To study tidal forces in a given spacetime, it is crucial to analyze the geodesics followed by massive test particles in that spacetime. Associated with the line element~\eqref{line}, we have the Lagrangian
\begin{equation}\label{Lag}
\mathcal{L}=\frac{1}{2}\,g_{\mu\nu}\,\dot{x}^{\mu}\dot{x}^{\nu},
\end{equation}
where the dot represents the derivative with respect to
the proper time $\tau$. For massive particles, we have $\mathcal{L}=-1/2$, and therefore
\begin{equation}
    -1=-A\dot{t}^2+\frac{1}{B}\dot{r}^2+r^2(\dot{\theta}^2+\sin^2\theta \dot{\phi}^2).
\end{equation}
For radial geodesics, we have $\dot{\theta}=\dot{\phi}=0$, and thus the equation above can be simplified as
\begin{equation}\label{eqcons1}
    -1=-\left(1-\frac{2M}{r}\right)\dot{t}^2+\frac{\dot{r}^2}{\left(1-\frac{2M}{r}\right)\left(1-\frac{a^2}{r^2}\right)}.
\end{equation}

From the Lagrangian~\eqref{Lag}, we can obtain the following conserved quantities
\begin{equation}
    E=-\frac{\partial \mathcal{L}}{\partial\dot{t}}=A\dot{t}, \quad L=\frac{\partial \mathcal{L}}{\partial\dot{\phi}}=r^2\sin^2\theta\dot{\phi}=0.
\end{equation}
The constants $L$ and $E$ represent the angular momentum and the energy of the particle per unit mass, respectively. Using the constants of motion, we can write the following energy-balance equation
\begin{equation}
    -1=-\frac{E^2}{A}+\frac{\dot{r}^2}{B}.
\end{equation}
From this relation, we can determine the energy of a particle released from rest at the radius $r=b$, which is given by
\begin{equation}
    E=\sqrt{A(r=b)}=\sqrt{1-\frac{2M}{b}}.
\end{equation}
The particle energy will be important when we study the tidal forces in the case of a test body falling radially into the black hole.
    
\section{TIDAL FORCES}\label{SEC:TF}
In a curved spacetime, two nearby geodesics that are initially parallel may approach or separate. The relative acceleration between these geodesics is related to the tidal forces of the spacetime under consideration. This relative acceleration is described by the geodesic deviation equation, given by \cite{dInverno:1992gxs}
\begin{equation}
    \frac{D^2\xi^\mu}{D\tau^2} = K^\mu_{\;\;\;\gamma}\; \xi^\gamma,
\end{equation}
where $\xi^\gamma$ is the infinitesimal displacement vector between nearby geodesics and $K^\mu_{\ \gamma}$ is the tidal tensor given in by
\begin{equation}
    K^\mu_{\ \gamma} = R^\mu_{\ \alpha\beta\gamma}v^{\alpha}v^\beta,
\end{equation}
where $R^{\mu}_{\ \alpha\beta\gamma}$ is the Riemann tensor and $v^\alpha$ is the timelike vector tangent to the geodesic. The components of the tidal tensor provide information about the tidal forces exerted by a given compact object on a test body.

Since tensor components do not directly correspond to physically measured quantities and may contain coordinate artifacts, it is standard to project the components of the displacement vector and the tidal tensor onto an orthonormal tetrad adapted to the observer's four-velocity. This tetrad defines the observer's local inertial frame, providing a better physical interpretation. This procedure yields coordinate-independent results, makes explicit the dependence of tidal forces on the observer's state of motion.

We consider a tetrad basis given by
\begin{equation}
\hat{e}^{\mu}{}_{\hat{a}}
=\left(\hat{e}^{\mu}{}_{\hat{0}},\ \hat{e}^{\mu}{}_{\hat{1}},\ \hat{e}^{\mu}{}_{\hat{2}},\ \hat{e}^{\mu}{}_{\hat{3}} \right),
\end{equation}
where indices with hat denote tetrad indices. The chosen tetrad basis must satisfy the orthonormality condition
\begin{equation}
\hat{e}^{\mu}{}_{\hat{a}}\,\hat{e}^{\nu}{}_{\hat{b}}\,g_{\mu\nu}
=\eta_{\hat{a}\hat{b}}.
\end{equation}
The vector $\hat{e}_{\hat{0}}^{\;\; \mu}$ is the four-velocity of the observer and $\{{\hat{e}_{\hat{1}}^{\;\;\mu}, \hat{e}_{\hat{2}}^{\;\;\mu},\hat{e}_{\hat{3}}^{\;\;\mu}}\}$ form an orthonormal basis for the spatial directions of the reference frame
attached to the observer. Here, $\eta_{\hat{a}\hat{b}}$ represent the components of the Minkowski metric

In this formalism, the components of the tidal tensor are expressed with respect to the tetrad basis according to
\begin{equation}
K^{\hat{a}}_{\,\,\hat{b}}=e^{\hat{a}}_{\,\,\mu}e^\nu_{\,\,\hat{b}}K^\mu_{\,\,\nu}\,,
\end{equation} 
while the components of the displacement vector are
\begin{equation}
\xi^{\hat{a}}=e^{\hat{a}}_{\,\,\mu}\xi^\mu\,.
\end{equation}

From this point on, we must choose an appropriate tetrad basis to describe the type of reference frame under consideration. Usually, three types of tetrad choices are employed: one associated with a static observer, another associated with a radially infalling observer, and finally one associated with an observer in circular motion. In this work, we focus on the first two cases.

\subsection{Static Observer}
We begin the study of tidal forces by considering the tetrad basis associated with a static observer. The tetrad basis corresponding to the line element~\eqref{line} is given by \cite{LimaJunior:2025uyj}
\begin{equation}\label{tetrada1}
\begin{split}
    \hat{e}_{\hat{0}}^{\;\; \mu}&= \left(\frac{1}{\sqrt{A}},\;0,\;0,\;0\right),\quad
    \hat{e}_{\hat{1}}^{\;\;\mu}= \left(0,\;\sqrt{B},\;0,\;0\right),\\
    \hat{e}_{\hat{2}}^{\;\;\mu}&= \left(0,\;0,\;\frac{1}{r},\;0\right),\quad
    \hat{e}_{\hat{3}}^{\;\;\mu}= \left(0,\;0,\;0,\;\frac{1}{r\sin\theta}\right).
\end{split}
\end{equation}

The tidal tensor associated with this tetrad basis is diagonal and can be written as
\begin{equation}
    K^{\hat{a}}_{\;\;\hat{b}} = \text{diag}(0,K_1,K_2,K_3),
\end{equation}
where
\begin{eqnarray}
    K_1 &=& -\frac{B A''}{2 A}-\frac{A' B'}{4 A}+\frac{B A'^2}{4 A^2},\\
    K_2 &=& K_3=- \frac{BA'}{2rA}.
\end{eqnarray}

For the Simpson--Visser spacetime, the components of the tidal force are given by
\begin{equation}\label{kstatic}
    K_1 = \frac{M(2r^2-3 a^2)}{r^5},\qquad
    K_2 =\frac{M \left(a^2-r^2\right)}{r^5}.
\end{equation}
It is straightforward to note that both the radial and angular components of the tidal forces vanish at
\begin{equation}\label{rA0est}
    r_0^R=\sqrt{\frac{3}{2}} a, \quad \mbox{and} \quad r_0^A=a
\end{equation}
respectively. This means that the angular component always vanishes at the wormhole throat. As for the radial component, depending on the value of $a$, the point at which $K_{1}=0$ can be located inside or outside the event horizon. More specifically, if $a>2\sqrt{2/3}\,M$, the radial component vanishes outside the event horizon. Since $r\geq a$, we can see that the regularization parameter reduces the intensity of both the radial and angular components; thus, the larger the value of this parameter, the weaker the tidal forces. In the case of the angular component, there is no sign inversion of the force, so $K_{2}=0$ is the smallest allowed value to the intensity of the angular component. For the radial component, however, a sign inversion occurs and $K_{1}$ becomes negative in the interval $a\leq r<\sqrt{3/2}\,a$.

By solving conditions $dK_{1}/dr=0$ and $dK_{2}/dr=0$, we can obtain the extrema of the tidal forces. At these points, we can also compute the second derivative to determine whether the extrema correspond to maxima or minima. In this way, we find that these points are given by
\begin{equation}
    r_{max}^R=\sqrt{\frac{5}{2}} a, \quad \mbox{and} \quad r_{min}^A=\sqrt{\frac{5}{3}}a.
\end{equation}
Therefore, we find that the radial component attains a maximum value given by
\begin{equation}
K_{1}^{\max}=\frac{8\sqrt{2/5}\,M}{25\,a^{3}},
\end{equation}
which can be located outside the event horizon if $a>2\sqrt{2/5}\,M$. 
On the other hand, the angular component attains a minimum value given by
\begin{equation}
K_{2}^{\min}=-\frac{6\sqrt{3/5}\,M}{25\,a^{3}},
\end{equation}
which can be located outside the event horizon if $a>2\sqrt{3/5}\,M$.

In this way, the differences introduced by the Simpson--Visser spacetime with respect to the Schwarzschild spacetime can be observable outside the event horizon, depending on the value of the wormhole throat radius.

While the static observer frame provides a valuable mathematical baseline for analyzing the components of the tidal tensor, its physical realizability is strictly limited. Maintaining a fixed radial coordinate close to the event horizon requires a divergent proper acceleration, making this scenario impossible for realistic compact objects. In actual astrophysical environments, processes such as tidal disruption events occur along free fall geodesics rather than static worldlines. Therefore, to accurately describe the realistic tidal forces experienced by a star plunging into the Simpson--Visser black bounce, it is necessary to evaluate the tidal tensor in the frame of a radially infalling observer, which we address in the following subsection.

\subsection{Radially infalling observer}
We now study the tidal forces experienced by an observer falling radially into the black hole. For this configuration, the tetrad basis associated with such an observer is given by \cite{Crispim:2025cql}
\begin{equation}\label{tetrada2}
\begin{split}
    \hat{e}_{\hat{0}}^{\;\; \mu}&= \left(\frac{E}{\sqrt{A}},-\;\sqrt{\frac{B}{A}}\sqrt{E^2-A},\;0,\;0\right),\\
    \hat{e}_{\hat{1}}^{\;\;\mu}&= \left(-\frac{\sqrt{E^2-A}}{A},\;E\sqrt{\frac{B}{A}},\;0,\;0\right),\\
    \hat{e}_{\hat{2}}^{\;\;\mu}&= \left(0,\;0,\;\frac{1}{r},\;0\right),\quad
    \hat{e}_{\hat{3}}^{\;\;\mu}= \left(0,\;0,\;0,\;\frac{1}{r\sin\theta}\right).
\end{split}
\end{equation}

The tidal tensor in this basis is given by
\begin{equation}
   \Bar{K}^{\hat{a}}_{\;\;\hat{b}} = \text{diag}(0,\Bar{K}_1,\Bar{K}_2,\Bar{K}_3),
\end{equation}
where
\begin{eqnarray}
    \Bar{K}_1 &=& -\frac{B A''}{2 A}-\frac{A' B'}{4 A}+\frac{B A'^2}{4 A^2},\\
    \Bar{K}_2 &=& \Bar{K}_3=\frac{\left(E^2-A\right)B'}{2rA}-\frac{E^2BA'}{2rA^2}.
\end{eqnarray}
It is interesting to note that, despite the more complicated tetrad basis, the radial component of the tidal force does not depend on the particle energy and remains the same as in the tetrad associated with a static observer, i.e. $\Bar{K}_1=K_{1}$. For the angular components, however, the situation is different: the angular component is more complicated and depends explicitly on the particle energy. Thus, depending on the choice of tetrad, at least some components of the tidal tensor are modified. It is also worth noting that, in spacetimes where $A=B$, the component $\Bar{K}_{2}$ reduces to the same value obtained in the static case; therefore, in spacetimes such as Schwarzschild or Reissner-Nordström, this choice of tetrad does not modify the physical results.

Since $\Bar{K}_{1}=K_{1}$, we focus on the differences arising from $\Bar{K}_{2}$, which is given by
\begin{equation}\label{k2rad}
    \Bar{K}_{2}=\frac{a^2 \left(\left(E^2-1\right) r+3 M\right)-M r^2}{r^5}.
\end{equation}
Once again, it becomes clear that there is a difference with respect to the static observer case, since the angular component of the tidal force now depends explicitly on the particle energy.

\begin{figure}
    \centering
    \includegraphics[width=1\linewidth]{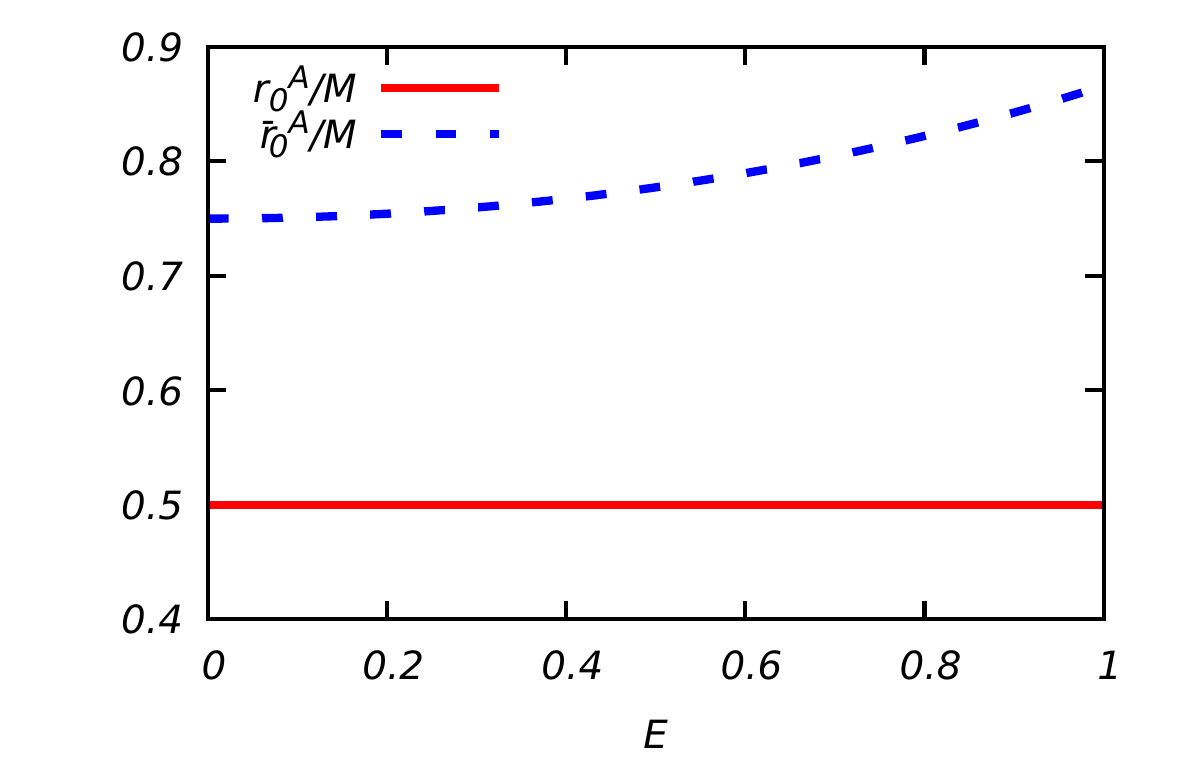}
    \caption{Location of the critical radius $r_0/M$, where the transverse (angular) tidal force vanishes for a radially infalling observer, as a function of the specific orbital energy $E/m$. The spacetime regularization parameter is fixed at $a = 0.5M$. Physically, $r_0$ marks the exact point where the transverse tidal force transitions between compression and stretching. This plot highlights that, unlike static observers who always experience this transition precisely at the throat ($r=a$), radially infalling observers encounter this transition increasingly earlier (farther outward from the throat) as their energy $E$ increases.
}
    \label{fig:r0A}
\end{figure}

To assess the modifications induced by the presence of energy on the tidal forces, we first determine the point at which the angular component of the tidal force vanishes. This point is given by
\begin{equation}
   \Bar{r}_0^A= \frac{a^2 \left(E^2-1\right)+a\sqrt{a^2 \left(E^2-1\right)^2+12 M^2}}{2 M}.
\end{equation}
Note that for a static observer, the angular component of the tidal forces vanishes precisely at $r = a$, Eq. \eqref{rA0est}, meaning this transition point is permanently located at the throat of the black bounce spacetime. On the other hand, for a radially infalling observer, this critical transition point depends on both the topological regularization parameter $a$ and the specific conserved energy $E$. In Figure~\ref{fig:r0A} we compare the radius at which the angular component of the tidal force vanishes energy $E$, fixing the geometric parameter at $a = 0.5M$. The curve clearly demonstrates that an increase in the infalling energy monotonically pushes the transition point further outward, away from the throat. Physically, this implies that faster-moving objects will experience a reversal in the lateral tidal forces significantly earlier in their plunge than slower objects, demonstrating a novel kinematic-geometric coupling inherent to the Simpson--Visser spacetime.

\begin{figure}
    \centering
    \includegraphics[width=1\linewidth]{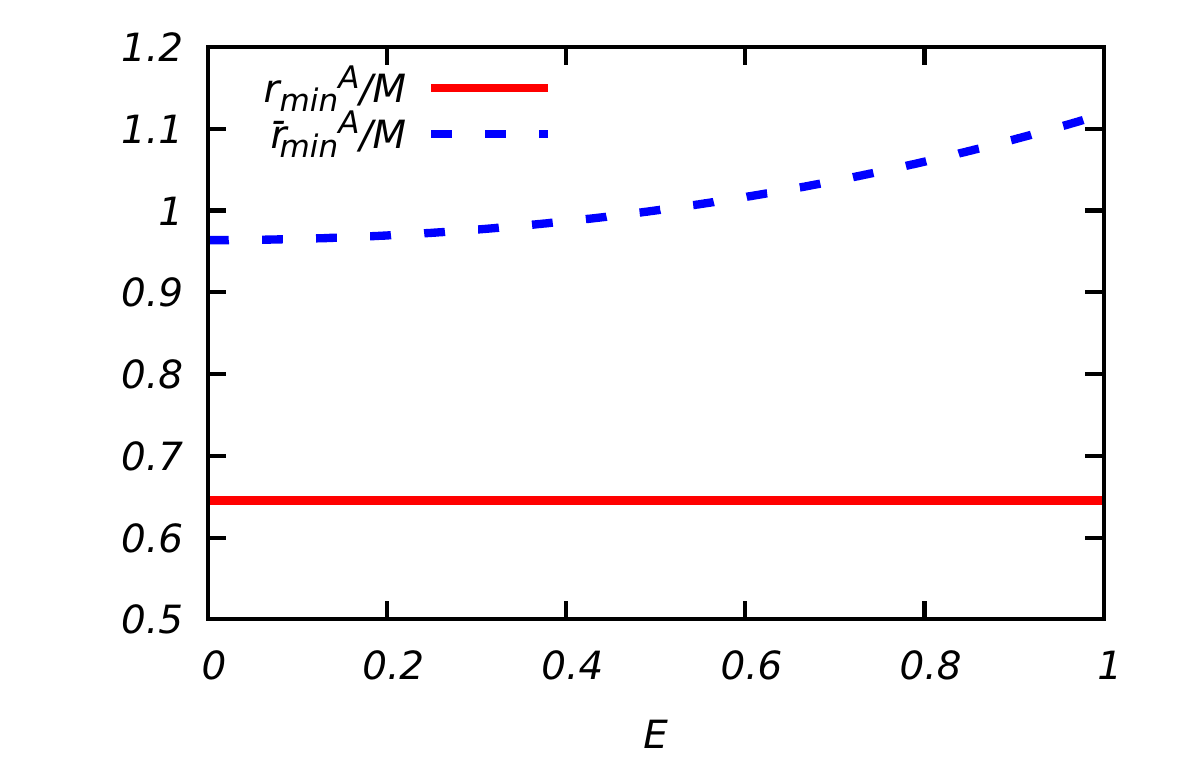}
    \caption{Comparison of the critical radius where the angular tidal force reaches its minimum value (point of maximum lateral compression), plotted as a function of the energy $E$. The regularization parameter is fixed at $a = 0.5M$. The solid red line represents the static observer (independent of $E$), while the dashed blue line corresponds to the radially infalling observer. Physically, this plot highlights that the infalling observer always encounters maximum tidal compression further away from the throat compared to the static observer, and this outward displacement increases monotonically as the infall energy grows.
}
    \label{fig:rminA}
\end{figure}

We now compute the exact point at which the angular component of the tidal force attains its minimum, which physically corresponds to the region of maximum lateral compression, which is given by
\begin{equation}
    \Bar{r}_{min}^{A}=\frac{2 a^2 \left(E^2-1\right)+\sqrt{4 a^4 \left(E^2-1\right)^2+45 a^2 M^2}}{3 M}.
\end{equation}
To clearly illustrate the kinematic effects on this extreme tidal stress, In Figure~\ref{fig:rminA} we compare the radius at which the angular component of the tidal force reaches its minimum as a function of the specific energy $E$, maintaining the spacetime geometry fixed with $a = 0.5M$. The solid red line indicates the location of peak compression for a static observer, which naturally remains constant with respect to $E$. The dashed blue curve represents the infalling observer. We clearly observe that, even as the energy of the test body tends to zero, the infalling observer experiences maximum lateral compression at a significantly larger radius than the static observer. Furthermore, this critical radius strictly increases as the energy increases. Physically, this demonstrates that the kinematics of the free fall effectively push the zone of peak transverse tidal deformation outward, further away from the throat.

For the radially infalling case, the zero and the minimum of the angular component, analogously to the static case, can occur outside the event horizon. In this configuration, the point at which the angular component vanishes is hidden inside the event horizon if
\begin{equation}
a<\frac{2M}{\sqrt{1+2E^{2}}},
\end{equation}
while the minimum point is hidden by the event horizon if
\begin{equation}
a<\frac{2\sqrt{3}\,M}{\sqrt{7+8E^{2}}}.
\end{equation}
In this way, we see that the tidal forces in the Simpson--Visser spacetime differ from those in the Schwarzschild spacetime and, depending on the size of the wormhole throat, these differences can be located outside the event horizon and thus be observable.

In the next section, we will study how these modifications in the tidal forces affect the Roche limit for neutron stars, white dwarfs, and for Sun-like stars.

\section{Tidal disruption in the Simpson--Visser spacetime}\label{SEC:Roche}
The Roche limit characterizes the critical distance at which an extended body held together solely by its own self-gravity can no longer withstand the tidal forces generated by an external gravitational field \cite{Gezari:2021bmb}. As an object approaches a sufficiently compact gravitational source, such as a dense star or a black hole, the spatial variation of the gravitational field across its finite size produces differential accelerations that tend to stretch the object along the radial direction and compress it along the transverse directions \cite{Hobson:2006se}. When these tidal forces exceed the internal gravitational binding (or other cohesive forces), the object undergoes tidal disruption, thereby defining the Roche limit \cite{Komossa:2015qya}. 

In the context of relativistic compact objects, a standard analytical approximation for the Roche limit can be formulated by comparing the radial component of the tidal force with the Newtonian internal binding force of the extended body. Assuming as a first-order approximation that tidal disruption is driven primarily by radial stretching and that the star remains a perfectly spherical, incompressible body up to the exact point of disruption, we can estimate the Roche radius, $r_{Roche}$, using \cite{Junior:2026upy}
\begin{equation}
\frac{1}{\xi^{\hat{1}}}\frac{D^2\xi^{\hat{1}}}{D\tau ^2}=K_1=\frac{M_{\star}}{R^3_{\star}}, \label{roche}
\end{equation}
where $R_{\star}$ and $M_{\star}$ denote the radius and mass of the star, respectively. We can rewrite the equation for the Roche limit as
\begin{equation}\label{roche2}
    \frac{M_{\star}r^5}{R_{\star}^3}-2Mr^2+3Ma^2=0.
\end{equation}
In this way, since the Roche radius depends on both the radius and the mass of the star subjected to an external gravitational field, different stellar models exhibit distinct Roche limits. Consequently, it is necessary to explicitly specify the physical cases under consideration. In this work, we analyze three different stellar models: neutron stars, white dwarfs, and Sun-like stars.

It is important to emphasize that Eq. \eqref{roche2} provides a preliminary algebraic estimate. This classical approximation inherently neglects the transverse compressive tidal forces ($K_2$ and $K_3$), internal stellar hydrodynamics, and the continuous ellipsoidal deformation of the star prior to disruption. To rigorously address these theoretical limitations and capture the full dynamical evolution, we complement this analytical baseline with a numerical Affine Standard Model in the following section.

In order to determine the Roche limit, our analysis focuses on the radial component of the tidal tensor, $K_1$. The physical rationale for this choice is that $K_1$ governs the longitudinal stretching force (often referred to as spaghettification) exerted along the axis connecting the star to the black bounce. It is precisely this stretching force that counteracts and ultimately overcomes the star's self-gravity, leading to its tidal disruption. In contrast, the angular components ($K_2$ and $K_3$) generally induce transverse compression, which does not primarily drive the tearing process that defines the standard Roche limit. Furthermore, it is crucial to highlight that the radial tidal component $K_1$ is purely invariant under the transformation between the static and the radially infalling observer frames, as demonstrated in the previous sections. Consequently, the disruption threshold and the calculated Roche radius remain mathematically identical for both cases. This guarantees that, despite the physical unrealizability of the static observer near the horizon, our derived condition for tidal disruption is completely robust and directly applicable to realistic astrophysical scenarios where the star follows a free-fall trajectory.

Throughout this work, we adopt \emph{natural units}, setting $c = G = 1$. Within this convention, the masses of black holes and stars acquire the same physical dimension as length. In contrast, in the International System of Units (SI), mass and length do not share the same unit of measurement. Therefore, astrophysical data must be appropriately converted into the unit system employed here in order to ensure a consistent comparison between physical quantities. In particular, a mass $M$ expressed in SI units is mapped into a length scale according to $M \;\longrightarrow\; \frac{G M}{c^2},$
which numerically yields
$1\,M_\odot \simeq 1.477~\mathrm{km}$ \cite{Shapiro:1983du}.

The mass and radius data for the stars considered in this work are listed in Table~\ref{tab:stellar_models}.

\begin{table}[ht]
\centering
\caption{Representative mass and radius values for the stellar models considered in this work \cite{Shapiro:1983du}. Masses are expressed in units of solar mass $M_\odot$, while radii are given in kilometers.}
\label{tab:stellar_models}
\begin{tabular}{lccc}
\hline\hline
Stellar model & $M_\star\,[M_\odot]$ & $M_\star\,[\mathrm{km}]$ & $R_\star\,[\mathrm{km}]$ \\
\hline
Neutron star  & $1.4$ & $2.07$ & $12$ \\
White dwarf   & $0.6$ & $0.89$ & $7.0\times10^{3}$ \\
Sun-like star & $1.0$ & $1.48$ & $6.96\times10^{5}$ \\
\hline\hline
\end{tabular}
\end{table}

For the Schwarzschild case, it is possible to solve Eq.~\eqref{roche} analytically. For the Simpson--Visser spacetime, however, this cannot be done analytically, and therefore we extract information about the Roche limit for this spacetime through graphical and numerical analysis. 

To determine in a general form the parameter regions where tidal disruption is observable, the Roche radius must lie outside the event horizon, i.e., $r_{\rm Roche} > r_h$. In the coordinate system adopted for the Simpson--Visser spacetime, the horizon is located at $r_h = 2M$. Although the Roche radius is obtained from the roots of the fifth-order polynomial in Eq. \eqref{roche2}, which does not admit a simple closed-form solution in radicals, the boundary separating observable tidal disruption events from direct capture can still be determined analytically. This boundary corresponds to the critical configuration where the disruption occurs exactly at the horizon, $r_{\rm Roche} = 2M$.
Substituting this condition into Eq.~\eqref{roche2} yields
\begin{equation}
\frac{32 M_{\star} M^4}{R_{\star}^3} - 8M^2 + 3 a^2 = 0.
\end{equation}
Solving for the regularization parameter gives the critical value
\begin{equation}
a_{\rm crit} =
\sqrt{\frac{8}{3}M^2 - \frac{32 M_{\star} M^4}{3R_{\star}^3}}.\label{acrit}
\end{equation}
This expression defines the boundary between two regimes: when $a < a_{\rm crit}$ the Roche radius lies outside the horizon and tidal disruption is observable, whereas for $a > a_{\rm crit}$ the star plunges into the compact object before being disrupted. For sufficiently large black hole masses the second term inside the square root dominates, making the expression negative and eliminating real solutions for $a_{\rm crit}$. In this regime, the Roche radius necessarily lies inside the horizon, implying that tidal disruption events are hidden from an asymptotic observer.
\subsection{Roche limit for neutron star}

\begin{figure*}[htb!]
    \centering
    \includegraphics[width=.5\linewidth]{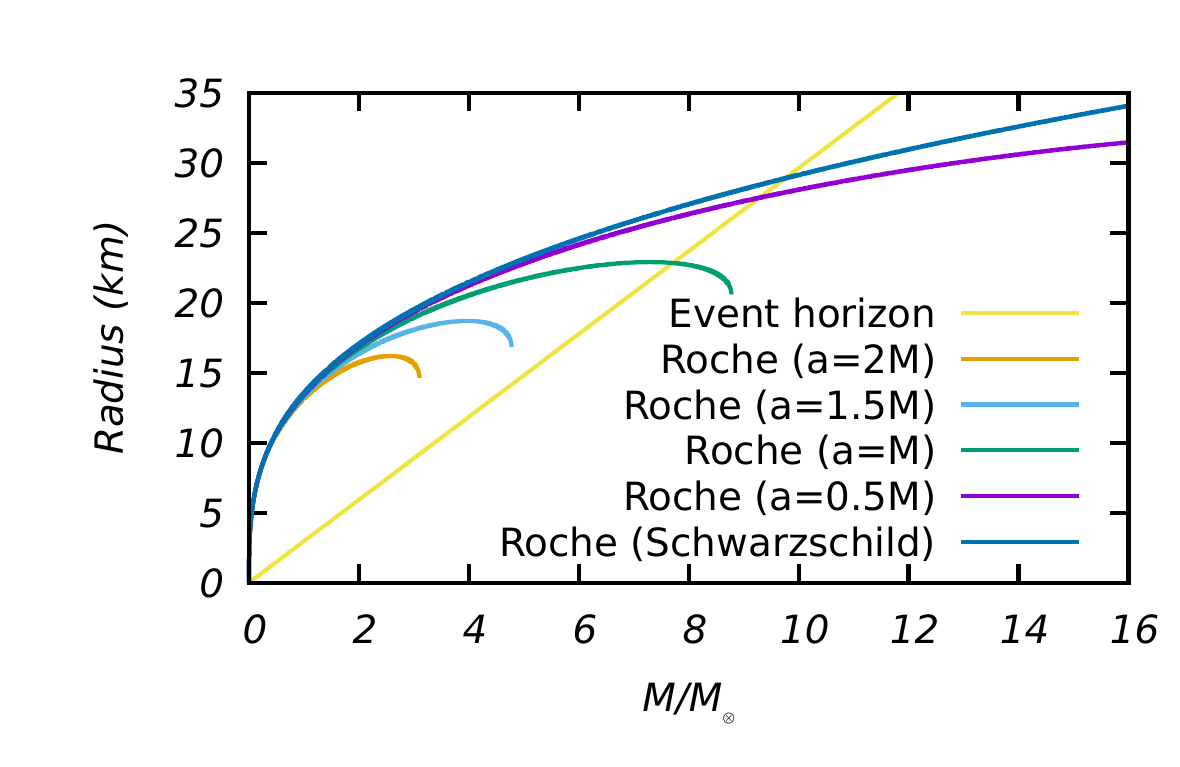}\hspace{-0.5cm}
    \includegraphics[width=.5\linewidth]{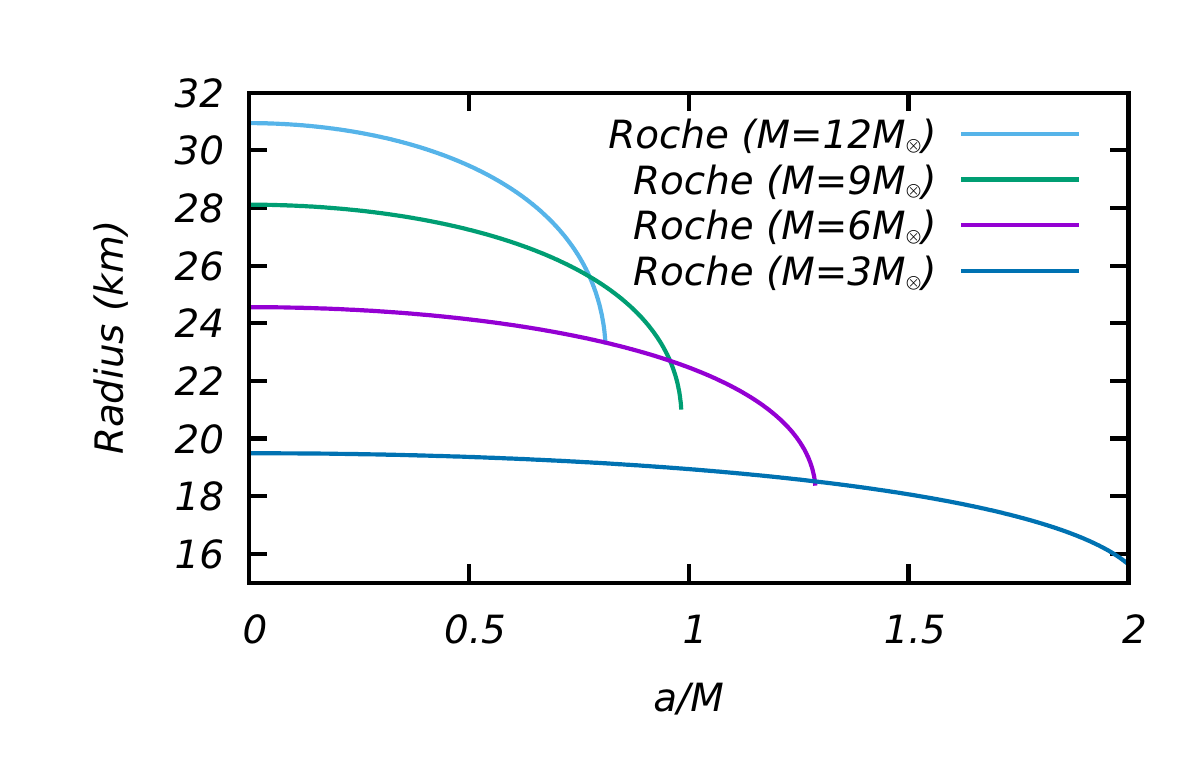}
    \caption{Roche radius for a neutron star (in km) evaluated within the Simpson--Visser geometry. Left panel: The Roche radius as a function of the black hole mass $M/M_\odot$ for different values of the regularization parameter $a/M$, compared with the event horizon (solid yellow line). Right panel: The Roche radius as a function of $a/M$ for fixed black hole masses. Physically, these plots demonstrate that geometric regularization diminishes the destructive power of tidal forces. As $a$ increases, the Roche radius shrinks, allowing the star to penetrate deeper into the gravitational well before disruption. Consequently, the likelihood of the neutron star being swallowed whole, without producing an observable tidal disruption event outside the horizon, significantly increases.}
    \label{fig:rocheNS}
\end{figure*}
We now analyze the behavior of the Roche limit in the Simpson-Visser spacetime by considering a typical neutron star. Using the data in Table \ref{tab:stellar_models}, we take the star to have a mass of approximately $1.4 M_\odot$ and a radius of about $12 km$. The corresponding results are shown in Figure \ref{fig:rocheNS}.In the left panel, we illustrate how the Roche radius varies with the black hole mass. A striking physical conclusion here is that the Simpson--Visser parameter $a$ imposes a maximum black hole mass for which a Roche radius even exists. If the regularization parameter $a$ is sufficiently large, the weakened tidal forces can no longer overcome the star's internal binding forces, meaning the star does not undergo disruption at all. For configurations where disruption still occurs, both panels confirm that the Roche radius strictly decreases as $a$ increases.

Astrophysically, this implies that for a fixed black hole mass, the disruption may occur outside the event horizon (an observable tidal disruption event), hidden inside the horizon, or not at all, depending entirely on the spacetime geometry. For instance, for a black hole with mass $M = 9 M_\odot$, the standard Schwarzschild case yields a Roche limit of $28.11 km$. Since this is larger than the event horizon radius ($26.64 km$), the disruption takes place outside and would be observable. However, introducing a Simpson-Visser parameter of $a = 0.7 M$ shrinks the Roche radius to $26.21 km$; consequently, the star crosses the horizon intact and disruption occurs only inside. If we further increase the parameter to $a = M$, the protective geometric effect ensures that disruption does not occur at all.

Finally, for small values of $a$ and sufficiently large black hole masses, the Roche radius naturally becomes hidden inside the event horizon, as expected from standard black hole astrophysics \cite{Xin:2025ipu,Mummery:2023meb}. In this regime, tidal forces evaluated near the horizon decrease rapidly with increasing black hole mass, ensuring that the disruption of neutron stars occurs strictly inside the horizon and remains unobservable.

For small values of $a$ and sufficiently large black hole masses, the Roche radius becomes hidden inside the event horizon, as expected \cite{Xin:2025ipu,Mummery:2023meb}. Tidal forces evaluated near the event horizon decrease rapidly as the black hole mass increases, so that the disruption of neutron stars occurs inside the event horizon and is therefore not observable.

\subsection{Roche limit for white dwarf}
As discussed in the previous subsection, for masses $M\approx 10M_{\odot}$ neutron stars undergo disruption only when it is hidden by the event horizon, if disruption occurs at all, even for small values of $a$. We now use Eq.~\eqref{roche2}, together with the data from Table~\ref{tab:stellar_models}, to verify whether this behavior also occurs for a white dwarf.
\begin{figure*}
    \centering
    \includegraphics[width=0.5\linewidth]{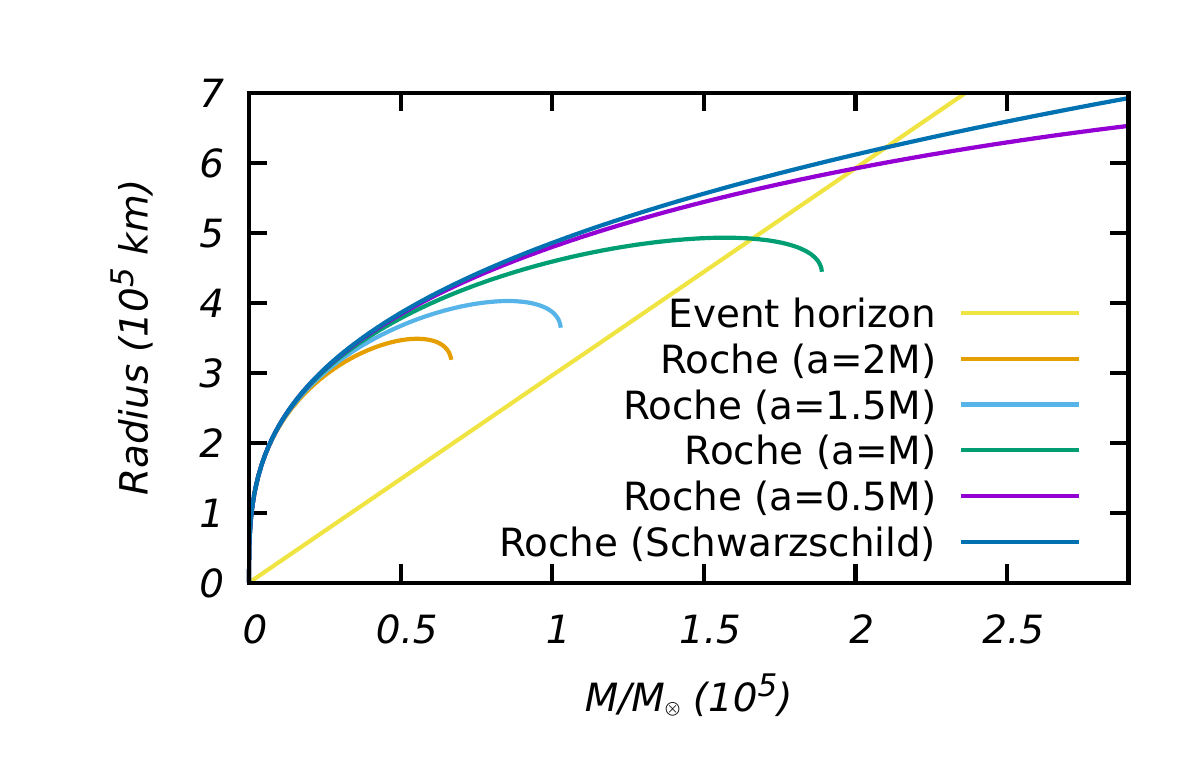}\hspace{-0.35cm}
    \includegraphics[width=0.5\linewidth]{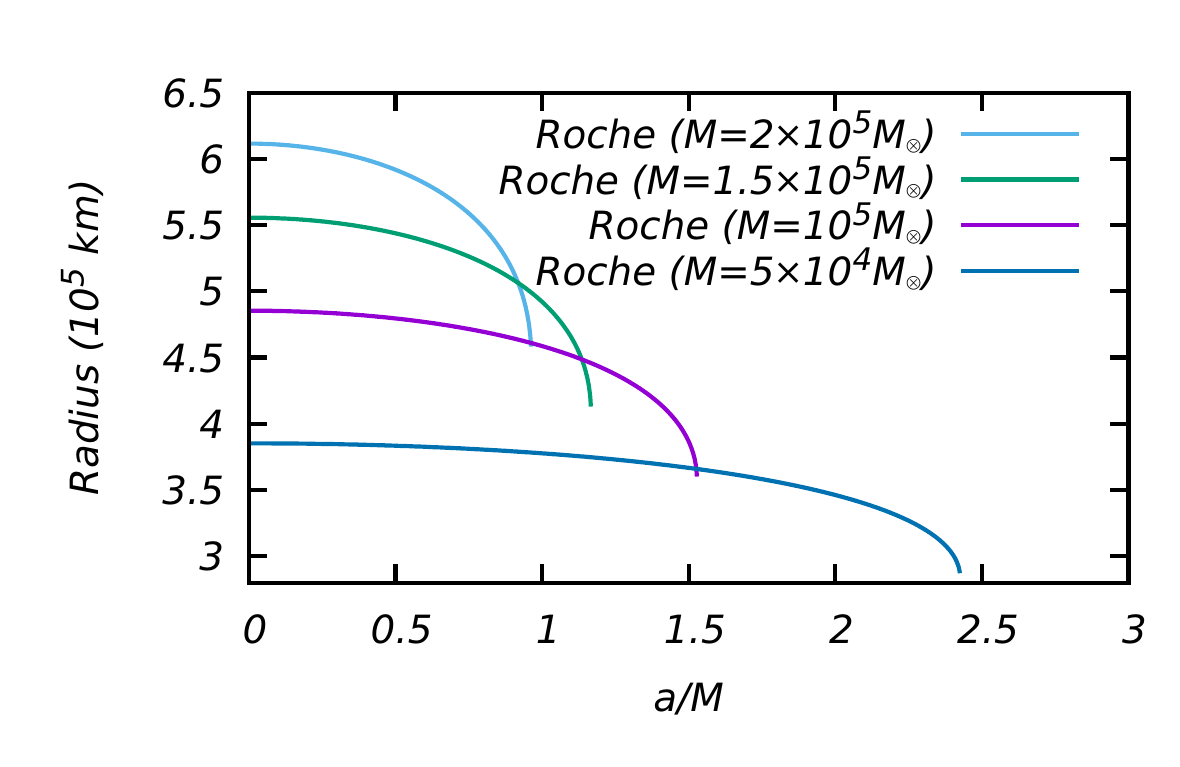}
    \caption{Roche radius for a typical white dwarf evaluated within the Simpson--Visser geometry. Left panel: The Roche radius (in units of $10^5$ $km$) as a function of the black hole mass $M/M_\odot$ (scaled by $10^5$) for varying regularization parameters $a/M$, compared with the event horizon (solid yellow line). Right panel: The monotonic decrease of the Roche radius as a function of $a/M$ for fixed black hole masses. Physically, these results extend the conclusions of the neutron star case to intermediate and supermassive black hole scales. The geometric regularization $a$ significantly suppresses tidal stretching.}
    \label{fig:roche_WD}
\end{figure*}

In Figure \ref{fig:roche_WD}, we extend our analysis to the Roche radius of white dwarfs, exploring the intermediate to supermassive black hole regime. In the left panel, we illustrate how the Roche radius varies with increasing black hole mass for selected values of the regularization parameter $a$. Consistent with the neutron star case, we observe a strong geometric suppression of tidal forces: larger values of $a$ drastically reduce the maximum black hole mass capable of producing a Roche radius. The right panel corroborates this trend for fixed black hole masses, demonstrating that more massive black holes permit disruption only for strictly smaller values of the regularization parameter.

For configurations where the combination of black hole mass and parameter $a$ still permits disruption, the increased black hole mass typically forces the Roche radius to be completely hidden inside the event horizon. To illustrate the astrophysical implications of this effect, we consider two well-known supermassive black holes: M87* with mass $M = 6.5 \times 10^9 M_\odot$ \cite{EventHorizonTelescope:2019dse}, and Sgr A* with mass $M = 4.297 \times 10^6 M_\odot$ \cite{EventHorizonTelescope:2022wkp}. For the case of M87*, a physical Roche radius exists only if the regularization parameter is extremely small, satisfying $a \lesssim 9.4 \times 10^{-4} M$. Even in this limit, the Roche radius ($r_{\text{Roche}} \approx 1.5 \times 10^7$ $km$) is entirely dwarfed by the massive event horizon ($r_+ \approx 1.92 \times 10^{10}$ $km$). Similarly, for Sgr A*, a Roche limit exists provided that $a \lesssim 1.2 \times 10^{-1} M$. Here, the Roche radius reaches $r_{\text{Roche}} \approx 1.39 \times 10^6$ $km$, which still remains strictly smaller than the event horizon radius of $r_+ \approx 1.27 \times 10^7$ $km$. Ultimately, in both astrophysical scenarios, the horizon completely encloses the disruption radius. The white dwarf is swallowed whole, and any potential tidal disruption process remains unobservable to asymptotic observers.

\subsection{Roche limit for Sun-like star}
As a final case, we now analyze the Roche limit for Sun-like stars and investigate whether the disruption process for this stellar model differs from the two previous cases.

\begin{figure*}
    \centering
    \includegraphics[width=0.5\linewidth]{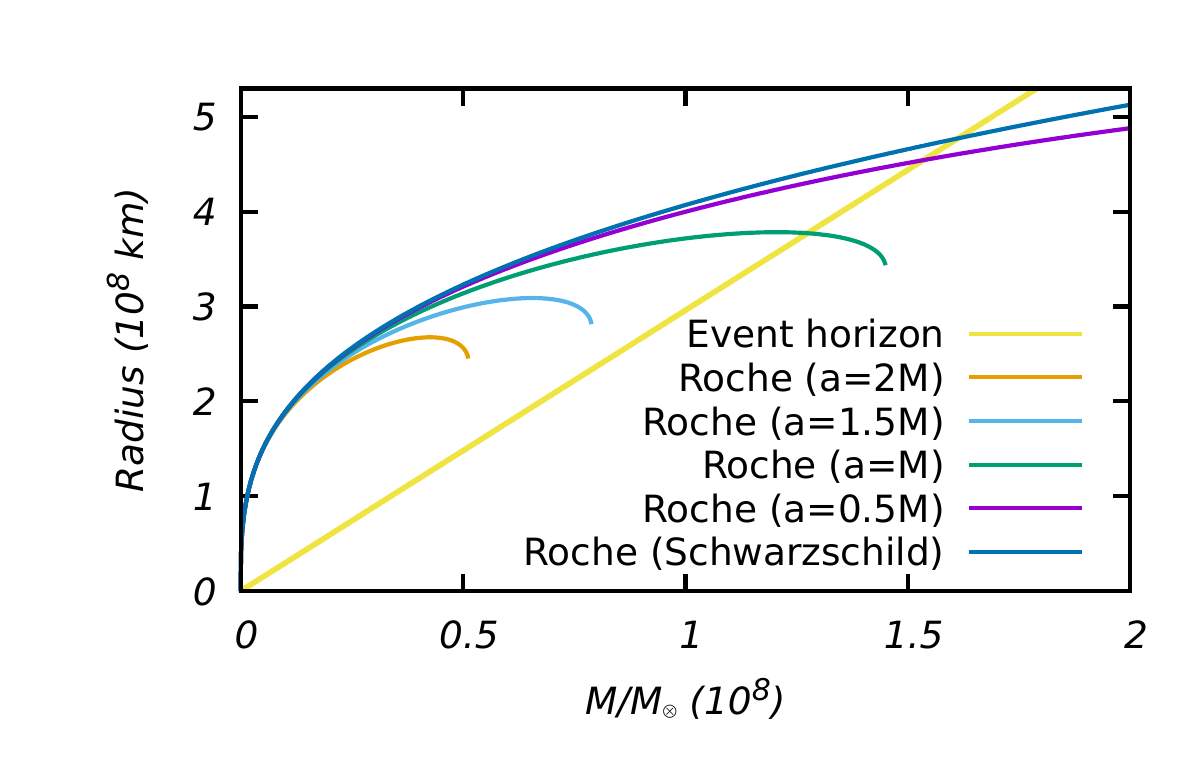}\hspace{-0.5cm}
    \includegraphics[width=0.5\linewidth]{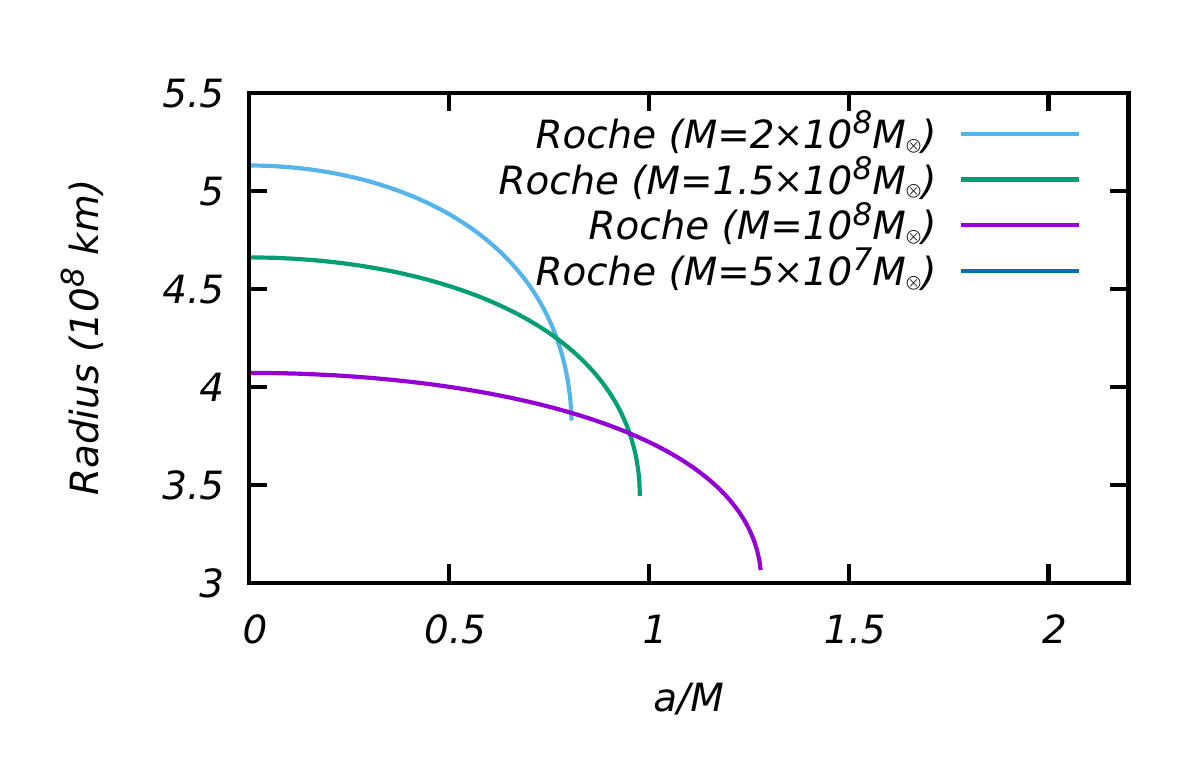}
    \caption{Roche radius for a Sun-like main-sequence star evaluated within the Simpson--Visser geometry. Left panel: The Roche radius (in units of $10^8$ $km$) as a function of the black hole mass $M/M_\odot$ (scaled by $10^8$) for varying regularization parameters $a/M$, compared with the event horizon (solid yellow line). Right panel: The monotonic decrease of the Roche radius as a function of $a/M$ for fixed supermassive black hole masses. Physically, this figure illustrates the tidal disruption limits for typical main-sequence stars encountering supermassive black holes. Consistent with denser stellar remnants, increasing the geometric regularization parameter $a$ strongly suppresses tidal forces.}
    \label{fig:roche_SL}
\end{figure*}

In Figure~\ref{fig:roche_SL}, we analyze the behavior of the Roche radius for Sun-like stars. In the left panel, we show how the Roche radius varies as the black hole mass is increased for selected values of the regularization parameter, and we observe that the larger the regularization parameter, the smaller the black hole mass for which the Roche radius exists. In the right panel, we fix the black hole mass and study how the Roche radius behaves as the regularization parameter is varied, finding the same trend: more massive black holes allow only smaller values of the regularization parameter.

For those cases in which the combination of black hole mass and regularization parameter permits disruption, more massive black holes have the Roche radius hidden inside the event horizon. For astrophysical black holes such as M87*, with mass $M=6.5\times10^{9}M_{\odot}$, Sun-like stars would have a Roche radius smaller than the event horizon radius, and thus the disruption process would be hidden by the event horizon.  For the case in which the black hole is as massive as M87*, a Roche limit exists only if the regularization parameter satisfies approximately $a\lesssim7.93\times10^{-2}M$. In this situation, the Roche radius is $r_{\text{Roche}}\approx1.21\times10^{9}\,\mathrm{km}$, while the event horizon radius is $r_{+}\approx1.92\times10^{10}\,\mathrm{km}$. Although the difference is much smaller than in the white-dwarf case, the disruption process would still be hidden inside the event horizon.

However, for the case of Sgr~A*, with mass $M=4.297\times10^{6}M_{\odot}$, the black hole mass is small enough that the Roche radius, for the values of $a$ considered, is larger than the event horizon radius. In this way, even in the case of a supermassive black hole such as Sgr~A*, the disruption process can be observable, allowing us to probe astrophysically how the Simpson--Visser regularization may influence the spacetime around the black hole. For black holes as massive as Sgr~A*, a Roche limit exists if the regularization parameter satisfies approximately $a\lesssim10.45M$, a value that is well above the upper bound for which an event horizon is still present. As an illustrative example within the horizon regime, we may consider $a=M$, for which the Roche radius is $r_{\text{Roche}}\approx1.42\times10^{8}\,\mathrm{km}$, while the event horizon radius is $r_{+}\approx1.27\times10^{7}\,\mathrm{km}$. In this case, the stellar disruption would occur outside the event horizon and would therefore be observable.

About the the possibility of no Roche limit, it is crucial to interpret the absence of a real solution for Eq. \eqref{roche2} with caution. This result is inherently model-dependent and strictly tied to the simplified Newtonian Roche limit approximation ($K_1 = M_*/R_*^3$). In a realistic astrophysical scenario, the lack of an analytical disruption threshold within this idealized framework does not guarantee absolute stellar stability. When full relativistic effects, the internal stellar equation of state, and complex hydrodynamic responses are taken into account, the star might still undergo disruption even in regions where the tidal field is weakened by the parameter $a$. Therefore, the conclusion that tidal disruption is completely avoided for large values of $a$ must be understood solely as a qualitative feature of this zeroth-order approximation.

From the expression for the critical parameter, Eq.~\eqref{acrit}, we can determine the corresponding relation for different stellar models. Using the representative stellar masses and radii listed in Table~I, we compute the critical parameter $a_{\rm crit}$ for each stellar type. The resulting expressions are summarized in Table~\ref{tab:acrit}. In this way, we can also understand why the Roche limit curves disappear for sufficiently large black hole masses, when the black hole mass becomes too large, the quantity inside the square root becomes negative, making $a_{\rm crit}$ imaginary. In this regime tidal disruption can no longer occur.
\begin{table}[h]
\centering
\begin{tabular}{c c}
\hline
Stellar model & $a_{\rm crit}$ \\
\hline
Neutron star &
$a_{\rm crit}=M\sqrt{\frac{8}{3}-0.01278\,M^2}$ \\
White dwarf &
$a_{\rm crit}=M\sqrt{\frac{8}{3}-2.77\times10^{-11}M^2}$ \\
Sun-like star &
$a_{\rm crit}=M\sqrt{\frac{8}{3}-4.68\times10^{-17}M^2}$ \\
\hline
\end{tabular}
\caption{Critical value of the parameter $a$ separating observable tidal disruption events from direct capture for different stellar models.}
\label{tab:acrit}
\end{table}

Before proceeding, it is crucial to explicitly clarify the physical interpretation and the domain of validity of the adopted Roche-limit condition, $K_1 = M_*/R_*^3$. As demonstrated in Sec. \ref{SEC:TF}, the radial tidal component $K_1$ is invariant under radial boosts due to the algebraic structure of spherically symmetric spacetimes. Therefore, Eq. \eqref{roche2} is geometrically valid for both static and radially infalling observers. However, despite this exact geometrical property, the disruption threshold itself remains a highly idealized Newtonian approximation. It represents an instantaneous criterion that assumes the star behaves as a rigid sphere of uniform density up to the exact point of disruption, relying on purely Newtonian self-gravity. For compact and strongly relativistic objects like neutron stars, their internal structure, equation of state (EoS), and binding energy cannot be adequately characterized by the simple mass-radius combination $M_*/R_*^3$.

Furthermore, while the radial stretching ($K_1=\bar K_1$) is independent of the observer's kinematic state (static or radial), the transverse compression components ($\bar K_2$ and $\bar K_3$) explicitly depend on the specific energy of the infalling trajectory. A physically consistent and realistic astrophysical treatment must account for the simultaneous action of all tidal components deforming the stellar fluid over proper time $\tau$. To address these limitations and capture the true dynamical nature of tidal disruption in strong gravitational fields, we must move beyond the instantaneous rigid-sphere approximation. In the following section, we introduce a dynamically consistent framework by coupling our full infalling tidal tensor to the affine stellar model.

\section{Dynamical Stellar Disruption: The Affine Model}
As highlighted in the previous section, while the radial disruption condition $\bar K_1 = M_*/R_*^3$ provides valuable geometric insight into the Simpson--Visser spacetime, it fundamentally describes an instantaneous and idealized threshold. In a realistic astrophysical scenario, a star does not disrupt at the exact moment this algebraic condition is met. Instead, tidal disruption is a continuous dynamical process. As the star plunges toward the black bounce, it is subjected to the simultaneous action of the invariant radial stretching ($\bar K_1$) and the energy-dependent transverse compressions ($\bar K_2$ and $\bar K_3$), which deform the stellar fluid over its proper time $\tau$.

To accurately capture this dynamical evolution and address the limitations of the rigid-sphere approximation, we adopt the affine stellar model originally formulated by Carter and Luminet \cite{1982Natur296211C,1983A&A12197C,1985MNRAS21257L}. In this framework, the star is modeled as a compressible, self-gravitating fluid ellipsoid rather than a rigid body. Its principal semi-axes $R_i$ (with $i=1,2,3$) are allowed to deform continuously as the star moves along its free-falling geodesic.

This dynamical approach allows us to fully utilize the complete tidal tensor derived for the radially infalling observer (Section III.B), explicitly incorporating the kinematic energy dependence of the transverse components. Throughout this analysis, we adopt geometric units where $G = c = 1$. By tracking the star along its proper time $\tau$, the evolution of the semi-axes is governed by a system of coupled ordinary differential equations:
\begin{equation}\frac{d^2 R_i}{d\tau^2} = - \bar K_i(\tau) R_i + a_{\text{grav},i} + a_{\text{press},i}, \label{eq:affine_general}
\end{equation}
where the first term on the right-hand side represents the external tidal driving force from the Simpson--Visser geometry, which depends on the black hole mass $M$ and the regularization parameter $a$. The terms $a_{\text{grav},i}$ and $a_{\text{press},i}$ represent the internal restoring accelerations due to the stellar self-gravity and the internal fluid pressure, respectively.

To solve Eq. \eqref{eq:affine_general}, we specify these internal accelerations for a self-gravitating polytropic fluid. The pressure acceleration is given by:\begin{equation}a_{\text{press}, i} = \frac{\hat{\Pi}}{M_* R_i} \left( \frac{R_*^3}{R_1 R_2 R_3} \right)^{\Gamma - 1} ,\end{equation}where $\Gamma$ is the adiabatic index characterizing the stiffness of the stellar material and $\hat{\Pi}$ is a constant related to the initial internal energy of the star, determined by the hydrostatic equilibrium condition at $r \to \infty$. The self-gravitational acceleration for the ellipsoidal configuration is expressed through the Dirichlet integral:\begin{equation}a_{\text{grav}, i} = - \frac{3 M_*}{2} R_i \int_0^\infty \frac{du}{(R_i^2 + u) \Delta(u)},\end{equation}where $\Delta(u) = \sqrt{(R_1^2+u)(R_2^2+u)(R_3^2+u)}$ accounts for the redistribution of the stellar mass $M_*$ as the geometry deviates from spherical symmetry.

We consider three distinct stellar models characterized by their compactness $\mathcal{C} = M_*/R_*$ and their respective adiabatic indices: a Sun-like star ($\Gamma=5/3$), a white dwarf ($\Gamma=5/3$ or $\Gamma=4/3$), and a neutron star ($\Gamma=3$). By integrating Eq. \eqref{eq:affine_general} from a safe initial distance where the star is in equilibrium ($R_1 = R_2 = R_3 = R_*$), we can dynamically track its deformation. In this framework, disruption is no longer defined by a simple algebraic equality, but rather by the dynamical instability of the stellar axes. Specifically, disruption occurs when the radial stretching $R_1$ undergoes unbounded growth ($R_1 \gg R_*$), signifying that the tidal work has overcome the star's self-binding energy. In cases where the tidal field is finite, as encountered in the regular regions of the Simpson--Visser geometry, this is numerically identified when the radial expansion velocity remains positive even as the tidal intensity peaks, leading to the eventual formation of a tidal stream.

Having established the theoretical framework of the Affine Standard Model, we now apply it to evaluate the dynamical deformation of our selected stellar models (neutron stars, white dwarfs, and Sun-like stars) as they plunge towards the Simpson-Visser black hole. In this dynamic approach, the star is not treated as a rigid sphere, but rather as an ellipsoid whose principal axes continuously deform under the influence of the varying external tidal tensor

\subsection{Dynamical Deformation of Neutron Stars}
\begin{figure*}[!htb]
    \centering
    \includegraphics[width=0.95\linewidth]{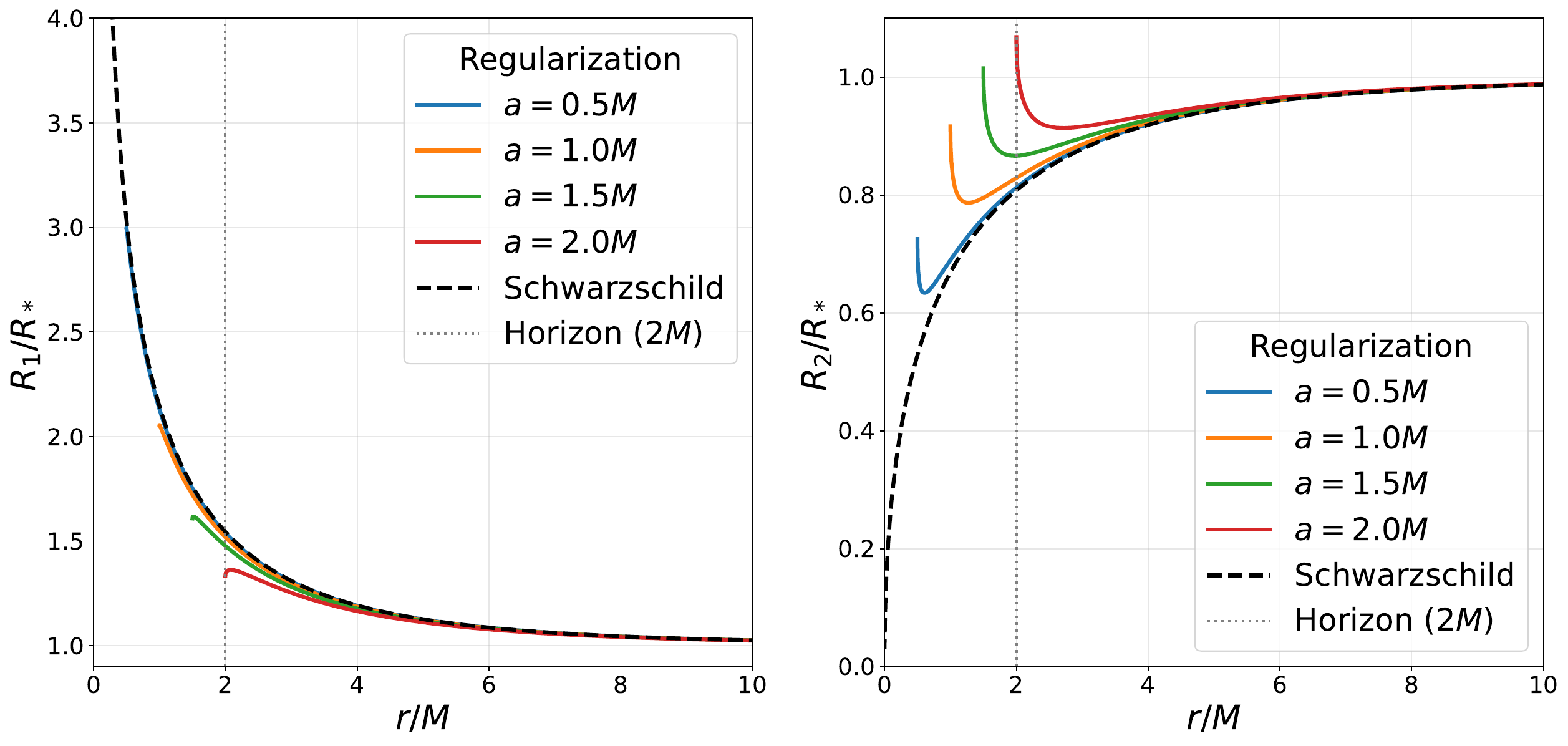}
    \includegraphics[width=0.95\linewidth]{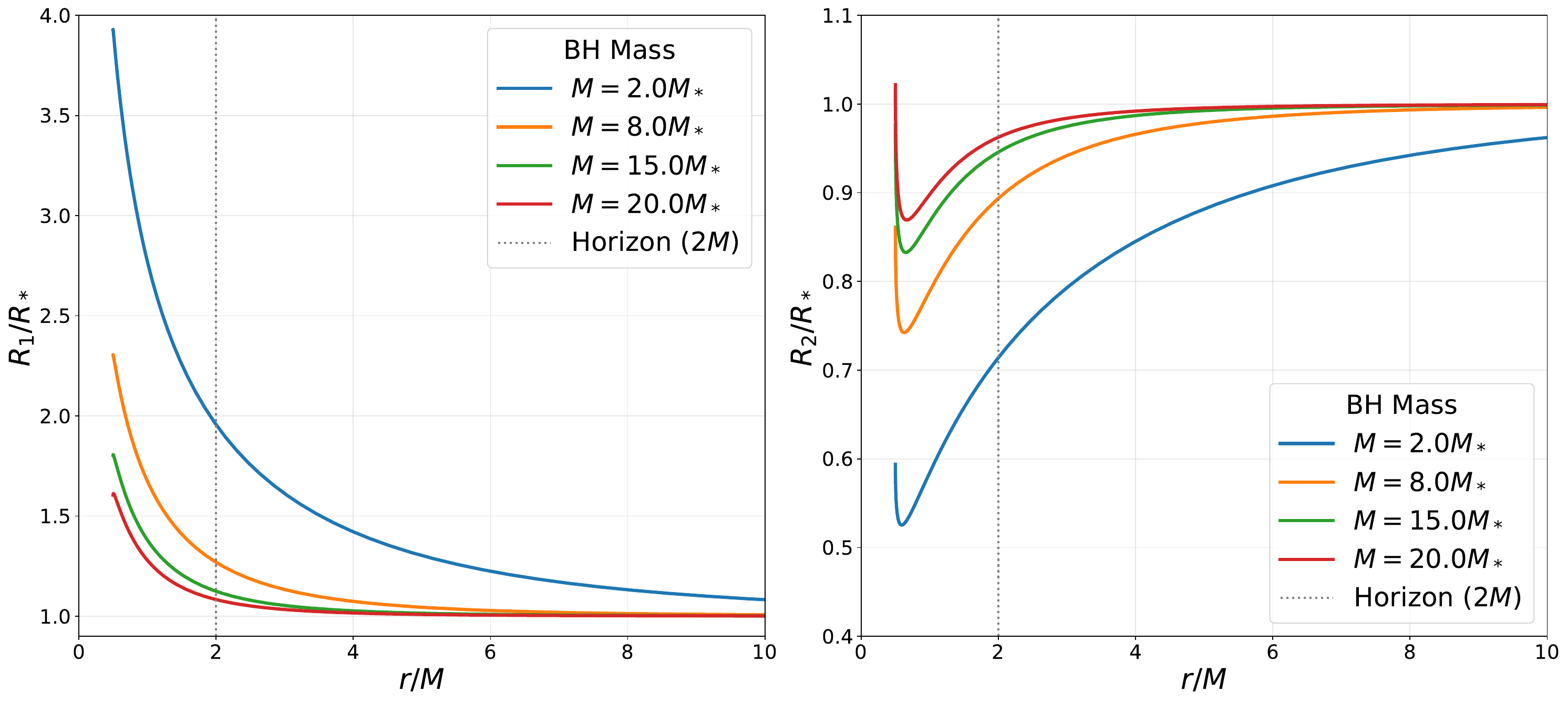}
    \caption{Dynamical evolution of the principal axes of a neutron star plunging into a Simpson--Visser black hole, evaluated using the Affine Model. The axes are normalized by the initial stellar radius $R_*$ and plotted against the dimensionless radial coordinate $r/M$. \textbf{Left panels:} Evolution of the radial axis $R_1/R_*$, representing the primary tidal elongation. \textbf{Right panels:} Evolution of the transverse axis $R_2/R_*$, representing transverse deformation. \textbf{Top panels:} The black hole mass is fixed at $M = 4M_*$ while the regularization parameter $a$ is varied. Increasing $a$ clearly dampens the tidal tensor: it drastically reduces the radial stretching of $R_1$, potentially preventing complete disruption, and introduces a distinctive behavior in $R_2$, where the axis undergoes a subsequent stretching phase following an initial compression. \textbf{Bottom panels:} The regularization parameter is fixed at $a = 0.5M$ while the black hole mass is varied. As the black hole mass increases, the tidal deformation of the star strictly decreases at a given dimensionless radius $r/M$.}
    \label{fig:AM_NS}
\end{figure*}
For highly relativistic and compact objects such as neutron stars, the internal self-gravity is exceptionally strong, dominating the external tidal field until the star reaches the immediate vicinity of the black hole. By employing the Affine Standard Model, we move beyond the rigid-sphere approximation and dynamically track the evolution of the star's principal axes as it approaches the Simpson--Visser black bounce.

The quantitative results of this dynamical evolution are presented in Fig. \ref{fig:AM_NS}, where we plot the normalized principal axes of the neutron star, $R_1/R_*$ (radial elongation) and $R_2/R_*$ (transverse deformation), as a function of the dimensionless radial coordinate $r/M$. The top panels of Fig. \ref{fig:AM_NS} illustrate the impact of the regularization parameter $a$ for a fixed black hole mass ($M = 4M_*$). As the star falls deeper into the gravitational well, the radial axis $R_1$ experiences significant stretching, while the transverse axis $R_2$ is initially compressed. However, the geometric suppression introduced by the Simpson--Visser spacetime becomes highly evident as $a$ increases. A larger regularization parameter severely dampens the eigenvalues of the tidal tensor, which drastically reduces the amplitude of the radial stretching in $R_1$. Furthermore, for sufficiently large values of $a$, the transverse axis $R_2$ exhibits a distinct and non-trivial behavior: after an initial phase of compression, the axis undergoes a subsequent stretching phase. This phenomenon is a direct dynamical signature of the modified tidal field in the regularized geometry, which can potentially prevent the complete triaxial disruption of the neutron star. In addition to the geometric shielding, our dynamical model accurately captures the mass-dependence of the tidal disruption process, as shown in the bottom panels of Fig. \ref{fig:AM_NS}. Here, the regularization parameter is fixed at $a = 0.5M$ while the black hole mass is varied. We observe that as the black hole mass increases, the amplitude of the stellar deformation at a given dimensionless radius $r/M$ strictly decreases. This dynamical behavior aligns perfectly with the standard astrophysical expectation: for a fixed dimensionless geometry, tidal forces evaluated near the event horizon are significantly weaker for more massive black holes. Consequently, supermassive black holes induce significantly less stretching on the infalling neutron star. This ensures that for supermassive configurations, the effective disruption point is pushed deeper into the gravitational well, often hiding the catastrophic event completely behind the horizon.

To provide a physical benchmark for disruption, we consider the threshold $R_1/R_* \gtrsim 2$, where the radial stretching typically overcomes the stellar self-gravity, leading to a tidal disruption. In this geometry, the radial coordinate $r$ is bounded by the throat located at $r=a$, and our dynamical plots are evolved until this geometric limit is reached.The top panels of Fig. \ref{fig:AM_NS} ($M = 4M_*$) reveal that for the Schwarzschild case ($a=0$), the neutron star experiences rapid deformation, exceeding the disruption threshold well before reaching the horizon. However, the introduction of the regularization parameter $a$ significantly alters this outcome. For $a=0.5M$, the stretching $R_1/R_*$ reaches a value near $3$ only as it approaches the throat. For the critical case $a=M$, the deformation at the event horizon is approximately $1.5$, reaching the value of $2$ only at the throat itself. Most notably, for the over-regularized case $a=1.5M$, the radial stretching remains remarkably low, staying below $1.6$ throughout the entire trajectory. This demonstrates that for sufficiently large values of $a$, the neutron star can reach the black bounce ($r=a$) structurally intact.In addition to the geometric shielding, the bottom panels ($a=0.5M$) demonstrate the mass-dependence of this process. Even with regularization, a black hole with $M=8M_*$ is still capable of inducing a deformation $R_1/R_* > 2$. However, as the black hole mass increases (e.g., $M=15M_*$ and $M=20M_*$), the amplitude of the stellar deformation at a given dimensionless radius $r/M$ strictly decreases, with the curves staying below the disruption threshold. This dynamical behavior aligns with the astrophysical expectation that tidal forces are weaker for more massive black holes at the horizon scale, pushing any potential disruption point deep into the gravitational well or beyond the horizon.

\subsection{Dynamical Deformation of White dwarf}
\begin{figure*}
    \centering
    \includegraphics[width=0.95\linewidth]{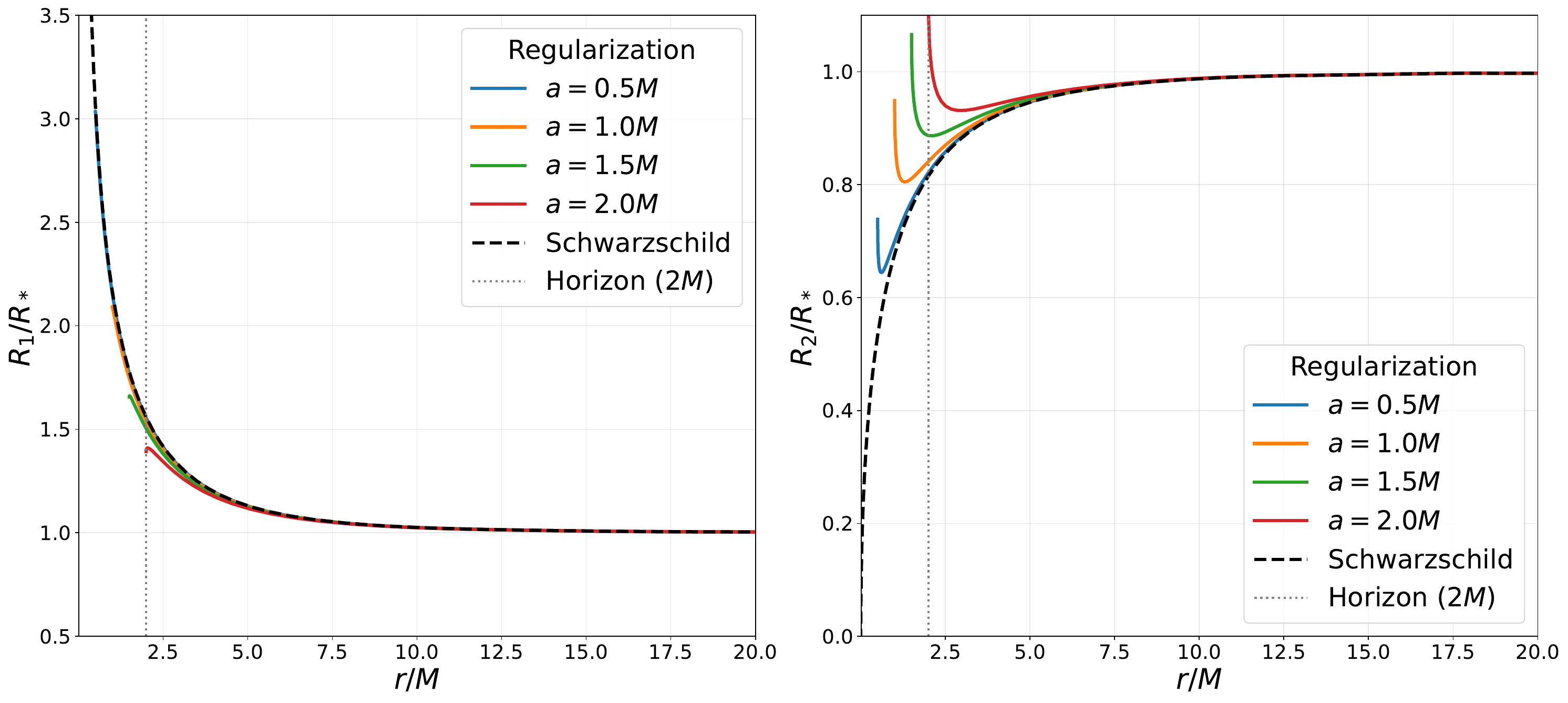}
    \includegraphics[width=0.95\linewidth]{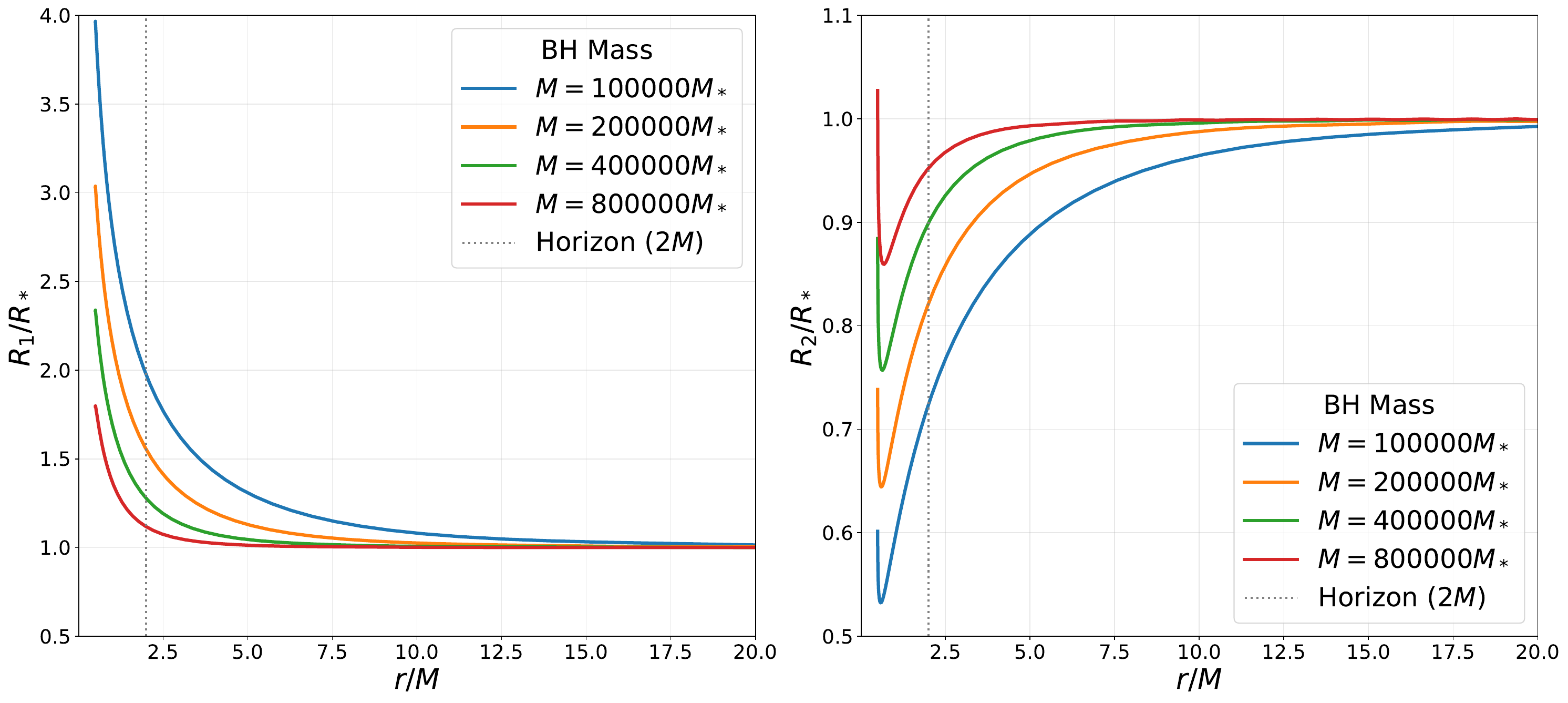}
   \caption{Dynamical evolution of the principal axes for a white dwarf in the Simpson--Visser spacetime, calculated via the Affine Model. \textbf{Top panels:} Fixed mass $M = 2 \times 10^5 M_*$ with varying $a$. We observe that $R_1$ (left) reaches the disruption threshold of 3 for small $a$, but is suppressed as $a$ increases. The transverse axis $R_2$ (right) shows a unique rebounding behavior, where it can exceed its initial radius ($R_2/R_* > 1$) near the throat for high regularization. \textbf{Bottom panels:} Fixed $a = 0.5M$ with varying black hole mass. Larger masses lead to reduced tidal stretching, shifting the potential disruption point deeper into the horizon or suppressing it entirely.}
    \label{fig:AM_WD}
\end{figure*}

White dwarfs represent an intermediate case of stellar compactness. Due to their lower binding energy compared to neutron stars, they are significantly more susceptible to tidal deformation. The dynamical evolution of these objects is presented in Fig. \ref{fig:AM_WD}, where we analyze their encounter with a supermassive black hole. In the top panels, we fix the black hole mass at $M = 2 \times 10^5 M_*$ and vary the regularization parameter $a$. Consistent with our previous findings, increasing $a$ provides a protective effect, drastically reducing the radial elongation $R_1$. For $a = 0.5M$, the star undergoes severe stretching, reaching $R_1/R_* \approx 3$ near the throat. However, for $a = M$, the deformation is moderated, staying just above the critical threshold of $2$. A remarkable feature observed in the right panel is the behavior of the transverse axis $R_2$. Unlike the monotonic compression expected in Schwarzschild gravity, the Simpson--Visser geometry allows $R_2$ to expand after an initial contraction. For larger values of $a$, this expansion can even result in $R_2/R_* > 1$, suggesting that the star becomes elongated in multiple directions as it approaches the black bounce. The bottom panels of Fig. \ref{fig:AM_WD} ($a = 0.5M$) show the influence of the black hole mass. For $M = 10^5 M_*$, the white dwarf is quickly disrupted, with $R_1/R_*$ exceeding 3. As the mass increases to $M = 2 \times 10^5 M_*$ and $M = 4 \times 10^5 M_*$, the tidal interaction becomes noticeably weaker at the same dimensionless radii. This confirms that even for white dwarfs, the ``safety'' of the star increases with the black hole mass in this coordinate scale, potentially allowing these objects to cross the event horizon of supermassive black holes without undergoing complete tidal disintegration.

\subsection{Dynamical Deformation of Sun-like stars}
\begin{figure*}
    \centering
    \includegraphics[width=0.95\linewidth]{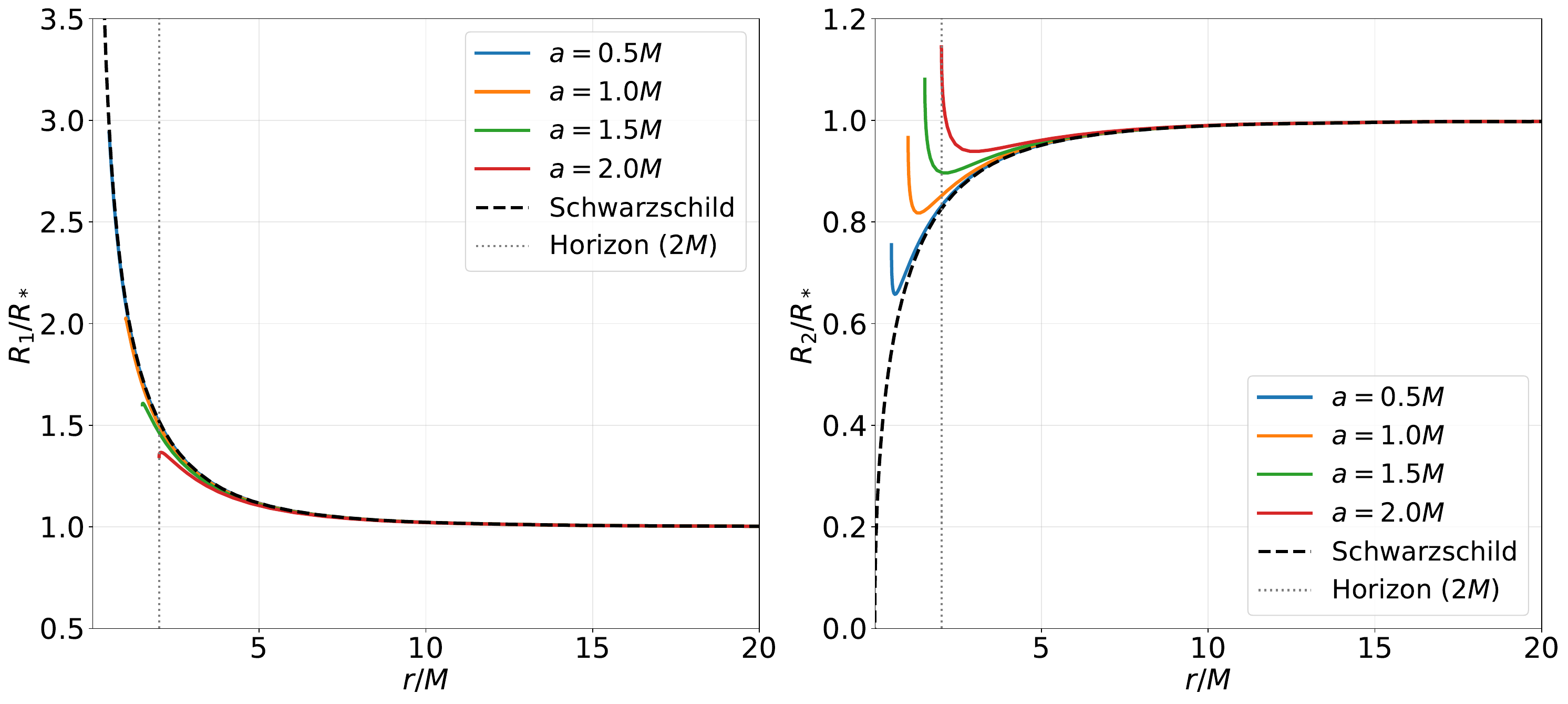}
    \includegraphics[width=0.95\linewidth]{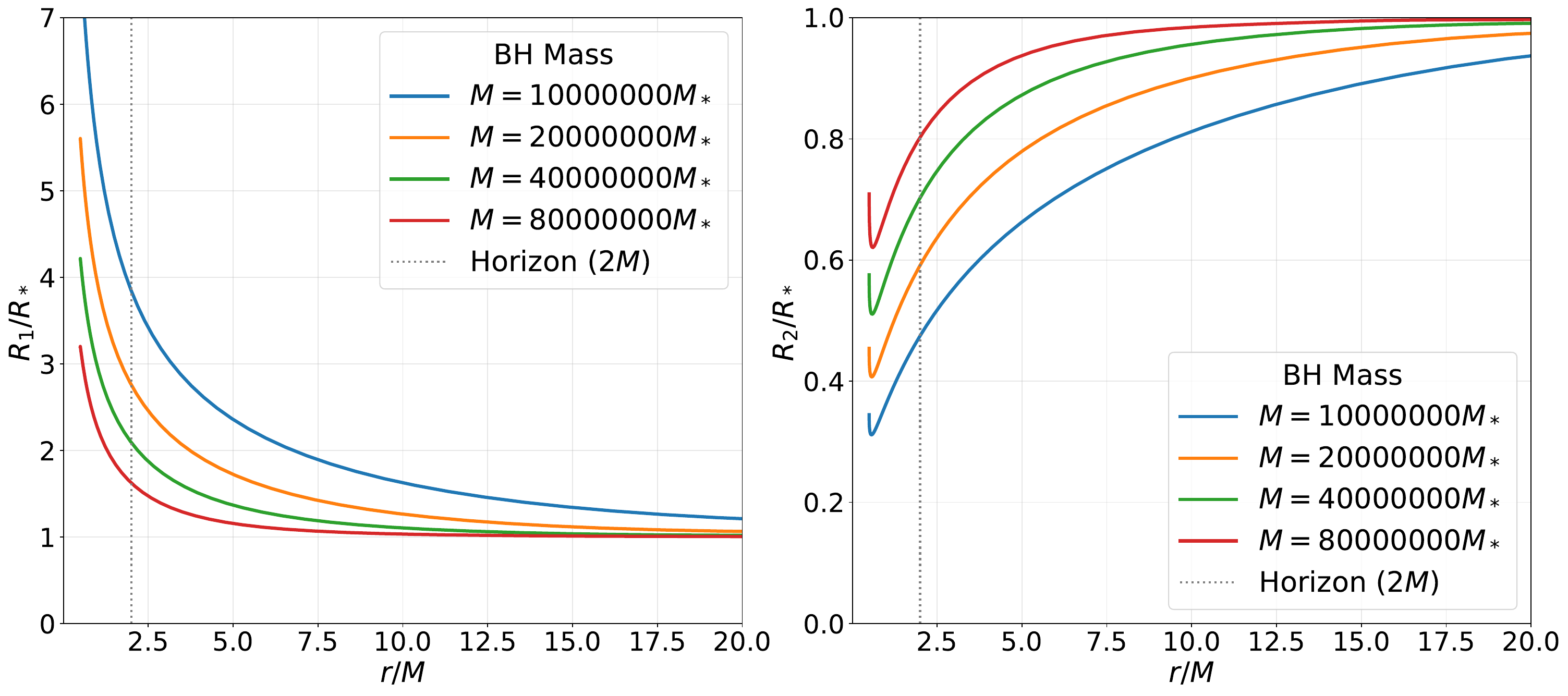}
    \caption{Dynamical evolution of the principal axes for a Sun-like star plunging into a Simpson--Visser spacetime, evaluated using the Affine Model. \textbf{Top panels:} Fixed black hole mass $M = 10^8 M_*$ with varying regularization parameter $a$. Due to the low stellar compactness, the radial axis $R_1$ (left) undergoes extreme stretching for low $a$, but is heavily suppressed for $a \geq 1.5M$. The transverse axis $R_2$ (right) exhibits a dramatic rebound effect for large $a$, significantly exceeding its initial radius ($R_2/R_* \gg 1$) and indicating strong transverse tidal stretching near the throat. \textbf{Bottom panels:} Fixed $a = 0.5M$ with varying black hole mass. Increasing the mass of the supermassive black hole significantly dampens the tidal deformation at a given $r/M$, delaying the disruption process.}
\label{fig:AM_Sun}
\end{figure*}
Main-sequence stars, such as the Sun, possess the lowest compactness and binding energy among the astrophysical objects considered in this study. Consequently, they are highly susceptible to external tidal fields and provide a sensitive probe for the modified tidal dynamics introduced by the Simpson--Visser geometry. The dynamical evolution of a Sun-like star falling into a supermassive black hole is presented in Fig. \ref{fig:AM_Sun}.

In the top panels of Fig. \ref{fig:AM_Sun}, the black hole mass is fixed at $M = 10^8 M_*$, and the regularization parameter $a$ is varied. For the standard Schwarzschild case ($a=0$) and low regularization ($a=0.5M$), the star undergoes extreme radial stretching, with $R_1/R_*$ rapidly exceeding the disruption threshold of $2$ and reaching values near $3$ at the throat. However, as $a$ increases, the protective geometric shielding becomes highly pronounced. For $a=M$, the maximum radial deformation is kept near $2$, and for $a=1.5M$, the star experiences almost negligible radial stretching ($R_1/R_* < 2$).The most striking feature of the regularized geometry in this low-density regime is the dynamical behavior of the transverse axis, $R_2$ (top right panel). Following an initial phase of compression, the transverse axis undergoes a dramatic rebound. For $a=1.5M$ and $a=2M$, this expansion far exceeds the initial stellar radius ($R_2/R_* > 1$). This indicates a complete inversion of the standard spaghettification process: in the highly regularized regime near the black bounce, the tidal tensor actively stretches the star in the transverse directions. Finally, the bottom panels ($a=0.5M$) illustrate the effect of varying the supermassive black hole mass from $M = 10^7 M_*$ to $M = 2 \times 10^7 M_*$. Even for easily disruptable main-sequence stars, an increase in the black hole mass systematically reduces the amplitude of both the radial stretching and the transverse compression at a given dimensionless radius $r/M$. This confirms that supermassive black holes induce significantly weaker tidal forces near their horizons, allowing even Sun-like stars to penetrate deeper into the gravitational well before the onset of catastrophic disruption.

Although the affine model focuses on the continuous dynamical deformation of the star rather than pinpointing a specific instantaneous disruption radius like the rigid sphere approximation, the results of both approaches are qualitatively consistent. In the dynamical framework, we observe that the radial deformation is heavily suppressed by increasing either the regularization parameter $a$ or the black hole mass $M$, eventually reaching a regime where the stretching is no longer sufficient to trigger a tidal disruption event. This dynamical suppression perfectly mirrors our findings from the static solid sphere analysis, where the increase of these same two parameters caused the classical Roche limit to vanish entirely, similarly indicating that the star avoids complete disruption.
\section{Summary and conclusion}\label{Sec:Conclusion}
In this work, we studied the tidal forces and the Roche limit for a black bounce spacetime described by the Simpson--Visser metric. This spacetime possesses several remarkable features, such as the presence of a wormhole throat hidden behind an event horizon, as well as its ability to mimic with high efficiency some characteristics of the Schwarzschild black hole.

In Section~\ref{SEC:spacetime}, we present the spacetime under consideration and discuss some of its main properties, such as the fact that event horizons exist only if $a\leq2M$, whereas for $a>2M$ the solution describes a two-way traversable wormhole. We also discuss that, depending on the coordinate system adopted, the radius of the unstable photon orbit differs from the Schwarzschild case, while the radius of the black hole shadow remains the same in both cases, provided that $a\leq3M$. In addition, we present the radial geodesics, which are essential to determine the energy of a particle depending on the position from which it is released.

When comparing these results with other well-known solutions, the distinctive features of the Simpson--Visser spacetime become evident. In the Schwarzschild geometry, tidal forces diverge at the singularity without ever changing sign. In the Reissner-Nordström case and various regular black holes, tidal forces exhibit maxima, minima, and a sign change, which are typically associated with the structure of a Cauchy horizon or a regular core at $r=0$. In the Simpson--Visser spacetime, however, the divergence is avoided because the radial coordinate is restricted to the throat at $r \ge a$. More notably, the Simpson--Visser model reproduces the maxima and minima of the tidal forces, as well as their sign inversion (transitioning from a stretching to a compressive regime), exclusively due to its black bounce topological structure, without requiring the presence of a Cauchy horizon.

Another result that fundamentally distinguishes spacetimes like the Simpson--Visser geometry is the behavior of the tidal tensor under the change of reference frames. In conventional spherically symmetric black hole solutions, the tidal force components are identical for both static and free-falling observers. Our analysis reveals that, in the Simpson--Visser spacetime, only the radial component of the tidal tensor remains invariant between frames. The angular components, conversely, acquire an explicit dependence on the energy ($E$) of the radially infalling observer. This breaking of symmetry between the reference frames demonstrates that the parameter $a$ affects the transverse dynamical coupling of the object in a way different from classical singular black holes.

In Section~\ref{SEC:TF}, we derived the tidal forces for the Simpson--Visser spacetime. We developed the formalism to study tidal forces through the analysis of the components of the tidal tensor appearing in the geodesic deviation equation. However, obtaining the tidal-tensor components alone is not sufficient to analyze tidal effects; it is also necessary to choose an appropriate local frame that characterizes the type of observer. We first considered the case of a static observer by selecting a suitable tetrad basis and projecting the tidal tensor onto this frame. We found that the tidal forces in this case do not diverge, since $r=0$ does not belong to the manifold and the radial coordinate is restricted to the interval $[a,\infty)$. We obtained that the angular component vanishes at the wormhole throat, $r=a$, while the radial component vanishes at $r=\sqrt{3/2}\,a$. The radial component attains a maximum value $K_{1}^{\max}=\frac{8\sqrt{2/5}\,M}{25\,a^{3}}$ at $r=\sqrt{5/2}\,a$, whereas the angular component attains a minimum value $K_{2}^{\min}=-\frac{6\sqrt{3/5}\,M}{25\,a^{3}}$ at $r=\sqrt{5/3}\,a$. We also studied the case of a radially infalling observer and found that the radial component of the tidal force coincides with the static case. In contrast, the angular component is modified and now depends explicitly on the energy. The radii at which the angular component vanishes or reaches its minimum are larger than in the static case and increase as the energy increases, as shown in Figures~\ref{fig:r0A} and~\ref{fig:rminA}.

In Section~\ref{SEC:Roche}, we used the radial component of the tidal forces to determine the Roche limit for three types of stars: neutron stars, white dwarfs, and Sun-like stars. For neutron stars, which are highly compact and therefore have a large mass concentrated within a relatively small radius, the Roche radius is typically smaller than the black hole horizon radius when the black hole mass is of the order of $10M_{\odot}$. Moreover, the regularization parameter further reduces the Roche radius. As a consequence, the Roche limit may be observable in the Schwarzschild case but not in the Simpson--Visser spacetime, depending on the value of the regularization parameter. For a fixed black hole mass, if $a$ is sufficiently large, no real solution for the Roche radius exists, implying that the star does not undergo tidal disruption. Only for relatively small black hole masses and sufficiently small values of $a$ can the Roche limit be located outside the event horizon. Black holes in this mass range may be observed astrophysically in coalescence events accompanied by gravitational-wave emission \cite{LIGOScientific:2025slb}.

For white dwarfs, which are less compact than neutron stars, the Roche limit can be observable outside the event horizon over a much wider mass range. However, for supermassive black holes such as Sgr~A* and M87*, this limit is not observable, since the Roche radius is smaller than the event horizon radius. As in the previous case, for a fixed mass there exists a maximum value of $a$ beyond which the Roche radius ceases to exist, implying the absence of tidal disruption. For Sgr~A* and M87*, these critical values of $a$ are of the order of $10^{-1}M$ and $10^{-4}M$, respectively.

Finally, we analyzed the Roche limit for Sun-like stars and found that supermassive black holes such as Sgr~A* possess an observable Roche limit, since in this case the Roche radius exceeds the event horizon radius. This allows tidal disruption to occur even for relatively large values of the regularization parameter, well above the bound required for the existence of an event horizon. For black holes with masses comparable to that of M87*, however, the Roche radius becomes hidden inside the event horizon, making stellar disruption unobservable, and the regularization parameter must be restricted to values of order $10^{-2}M$ for disruption to occur.

In general, in all cases the regularization parameter decreases the Roche radius as $a$ increases, until a point is reached at which no real Roche radius exists and tidal disruption no longer occurs. Our analytical approximation suggests that the Simpson--Visser geometry may mitigate tidal forces, potentially hiding or avoiding disruption, though full hydrodynamic models are required to confirm stellar survivability.

To provide a more realistic and continuous description of the tidal disruption process, we advanced beyond the static rigid-sphere approximation by implementing the Affine Model. By dynamically tracking the ellipsoidal deformation of neutron stars, white dwarfs, and Sun-like stars plunging into a Simpson--Visser black bounce, we demonstrated that the regularization parameter $a$ exerts a geometric shielding effect. Across all stellar compactness regimes, an increase in $a$ drastically dampens the eigenvalues of the tidal tensor, severely suppressing radial elongation and effectively shielding the star from complete disruption. A particularly striking signature of this regularized geometry is the dynamic behavior of the transverse axes. In contrast to the monotonic compression seen in classical spaghettification, the transverse axes can experience a rebound, stretching well beyond their initial radii as the star approaches the black bounce. Furthermore, our dynamical evolution verified that increasing the black hole mass consistently reduces the overall deformation amplitude. Ultimately, this dynamical framework fully corroborates our static Roche limit findings: both the regularization parameter and the black hole mass act in tandem to protect the infalling star, pushing the onset of disruption deeper into the gravitational well or averting it entirely.

In future work, we plan to investigate whether this behavior of the Roche limit persists for other black bounce models. As a natural extension of this work, we plan to investigate the tidal disruption limit in a rotating Simpson--Visser spacetime, since astrophysical black holes are not static, in order to determine how rotation influences the tidal forces and the Roche limit. Based on the Kerr geometry, where analysis is often simplified using axial trajectories or the ZAMO frame, rotation generally decreases tidal force intensity. This reduction implies that stars could potentially approach closer to the black hole before reaching the Roche limit or suffer less deformation at a given distance. Additionally, since the spin parameter can induce a sign inversion in the tidal components,similar to our regularization parameter $a$, we expect this transition from stretching to compression to be a persistent and potentially amplified feature in the rotating black bounce model.

\section*{Acknowledgements}

M.S. thanks Conselho Nacional de Desenvolvimento Cient\'ifico e Tecnol\'ogico - CNPq, Brazil, CNPQ/PDE 200218/2025-5, for financial support. The paper is also supported by the Spanish project PID2024-157196NB-I00 funded by MICIU/AEI/10.13039/501100011033.

\bibliography{ref.bib}

\end{document}